\input ./style/arxiv-general.cfg
\documentclass[MSNbibl,nameyear,dvips]{arxstspdf}
\makeatletter
   \@ifpackageloaded{graphicx}{}{\usepackage{graphicx}}
\makeatother
\usepackage{flushend}
\usepackage{stfloats}

\volume{30}
\issue{4}
\pubyear{2015}
\firstpage{468}
\lastpage{484}
\doi{10.1214/15-STS524}% Updated by VTEXPTS2LaTeX.exe, 07.08.2015 09:30
\docsubty{FLA}

\makeatletter
\renewcommand{\mid}{|}
\newcommand{\SRVF}{\mathrm{SRVF}}
\newcommand{\lleft}{\left}
\newcommand{\rrvert}{\vert}
\newcommand{\rright}{\right}
\newcommand{\rrVert}{\Vert}
\newcommand{\llvert}{\vert}
\newcommand{\llVert}{\Vert}
\renewcommand{\citep}[1]{(\citeauthor{#1}, \citeyear{#1})}
\newcommand{\real}{\mathbb{R}}
\newcommand{\sign}{\operatorname{sgn}}
\def\argmin{\operatorname{argmin}}
\makeatother

\begin{document}
\begin{frontmatter}

\title{Functional Data Analysis of Amplitude and Phase Variation}%
%\thanksref{T1}
% kai straipsnis turi susijusiu diskusiju ir rejoinder'iu
%\relateddois{T1}{Discussed in \relateddoi{d}{10.1214/00-STSXXX}...;
%rejoinder at \relateddoi{r}{10.1214/00-STSXXXX}.}
\runtitle{Amplitude--Phase Variation}
%\pdftitle{}

\begin{aug}
% Corresponding author: J. Marron - marron@email.unc.edu% Updated by
%VTEXPTS2LaTeX.exe, 07.08.2015 09:30
\author[A]{\fnms{J. S.}~\snm{Marron}\corref{}\corref{}\ead[label=e1]{marron@unc.edu}},
\author[B]{\fnms{James O.}~\snm{Ramsay}\ead[label=e2]{ramsay@psych.mcgill.ca}},
\author[C]{\fnms{Laura M.}~\snm{Sangalli}\ead[label=e3]{laura.sangalli@polimi.it}}
\and
\author[D]{\fnms{Anuj}~\snm{Srivastava}\ead[label=e4]{anuj@stat.fsu}}
\runauthor{Marron, Ramsay, Sangalli and Srivastava}
%\pdfauthor{}

\affiliation{University of North Carolina,
McGill University,
Politecnico di Milano,
Florida State University}

\address[A]{J.~S. Marron is Professor,
Department of Statistics and Operation Research, University of North Carolina,
318 Hanes Hall, CB\# 3260, Chapel Hill, North Carolina 27599-3260, USA
\printead{e1}.}
\address[B]{James~O.~Ramsay is Professor Emeritus,
Department of Psychology, McGill University, 1205 Dr Penfield Avenue Montreal,
QQuebec H3A 1B1, Canada \printead{e2}.}
\address[C]{Laura~M.~Sangalli is Associate Professor,
MOX, Dipartimento di Matematica, Politecnico di Milano, Piazza L. da Vinci 32, 20133 Milano,
Italy \printead{e3}.}
\address[D]{Anuj~Srivastava is Professor,
Department of Statistics, Florida State University, Tallahassee, Florida 32306, USA
\printead{e4}.}

\end{aug}

% ABSTRACT
%
\begin{abstract}
The abundance of functional observations in scientific endeavors
has led to a significant development in tools for functional data
analysis (FDA). This kind of data comes with several challenges:
infinite-dimensionality of function spaces, observation noise, and so on.
\mbox{However}, there is another
interesting phenomena that creates problems in FDA.
The functional data often comes with lateral displacements/deformations
in curves,
a phenomenon which is different from the height or amplitude
variability and is
termed \textit{phase variation}. The
presence of phase variability artificially often inflates data
variance, blurs underlying data structures, and distorts
principal components.
While the separation and/or removal of phase from amplitude data is
desirable, this is a difficult problem. In particular, a
commonly used alignment procedure,
based on minimizing the $\mathbb{L}^2$ norm between functions, does
not provide satisfactory results.
In this paper we motivate the importance of dealing with the phase
variability and summarize
several current ideas for separating phase and amplitude components.
These approaches differ in the following: (1) the definition and
mathematical representation of phase variability, (2) the objective
functions that
are used in functional data alignment, and (3) the algorithmic tools
for solving estimation/optimization problems.
We use simple examples to illustrate various approaches and to provide
useful contrast between them.
\end{abstract}

% KEYWORDS
\end{frontmatter}

%
%%------------------------------------------------------------------------------------------------
%%------------------------------------------------------------------------------------------------

%%%%%%%%%%%%%%%%%%%%%%%%%%%%%%%%%%%%%%%%%%%%%%%%%%%%%%%%%%%%%%%%%%%%%%%%%%%%%%%%%%%%%%%%%%%%%%%%%%%%%%
%f1 #&#
%
\begin{figure*}

\includegraphics{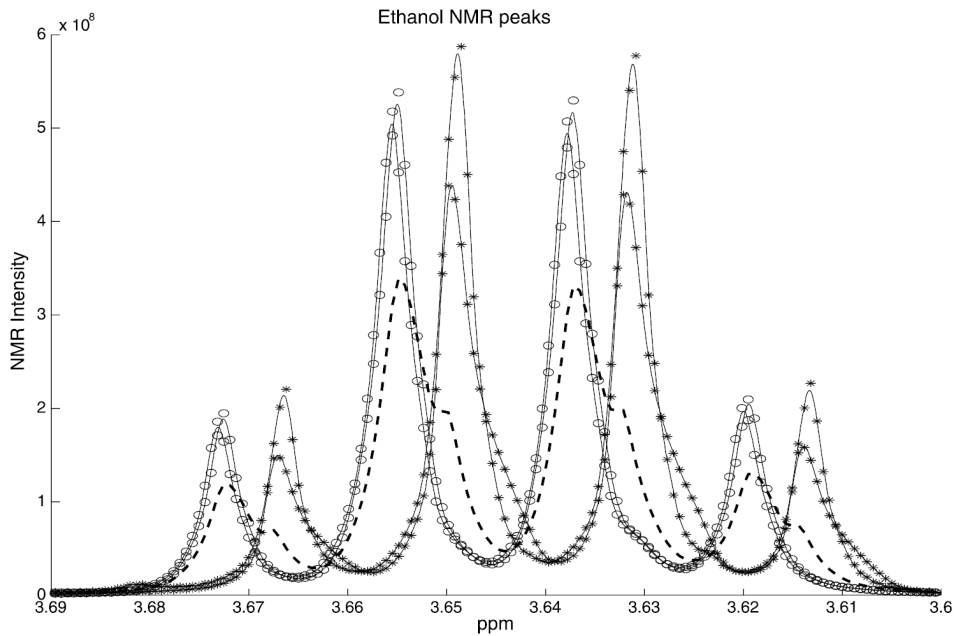}

\caption{Circles correspond to intensities over an ethanol region of
the NMR spectrum for two typical red wines, and asterisks indicate a
white and
a ros\'{e}. The light solid lines are smooth fits of the data using
order 6 B-spline basis functions with a knot at every sampling and a
light penalty ($\lambda= 10^4$) on the fourth derivative. The heavy
dashed line is the mean intensity across 31 reds, 7 whites, and 2 ros\'{e}s.
The mean has a far different shape from the quite similar shape of all
the data curves, due to registration issues.}
\label{ethanol}\label{fig1}
\end{figure*}
%
%%%%%%%%%%%%%%%%%%%%%%%%%%%%%%%%%%%%%%%%%%%%%%%%%%%%%%%%%%%%%%%%%%%%%%%%%%%%%%%%%%%%%%%%%%%%%%%%%%%%%%

%s1 #&#
\section{Introduction}\label{sec1}
\label{secintro}

%s1.1 #&#
\subsection{A First Look at Phase Variation in Functional Data}\label{sec11}
\label{subsecfirstphase}

Experimental units of data that are distributed over lines and areas,
known as functional data, are best represented as curves and surfaces,
respectively; and we expect that these will vary in height over any
particular point. But we often notice that the continuous substrate of
the data seems itself to be transformable, and that these
transformations vary across functional observations.

Figure~\ref{ethanol} displays four peaks for each of four samples of
wines in the part of the nuclear magnetic resonance (NMR) spectrum
corresponding to ethanol. Two of these wines are red, one is white, and
one is a ros\'{e}. We notice that most of the variation across these
four samples is due to the peaks of the white and ros\'{e} wines being
displaced to the right relative to those for the red wines. It is known
that the pH level in a solution has this effect on the location of the
couplets, triplets, and m-tuplets that NMR generates; and also that red
wines have pH's from 3.3 to 3.5, while white pH's are in the range
3.0--3.3. Moreover, the effects of pH and other factors are known to
vary from one location in the spectrum to another, with displacements
in opposing directions not being unusual.

The functional data analysis (FDA) literature refers to lateral
displacements in curve features as \emph{phase variation}, as opposed
to \emph{amplitude variation} in curve height. As in music, we imagine
that time can be compressed or stretched over different intervals in a
single performance. Consequently, we distinguish between measured clock
time and related but different time scales. Relative to human growth
time, for example, puberty for girls occurs on
average at the age of 11.7 years, but hormonal and other physiological
factors shift this age forward and backward to the variable clock times
that parents actually see.

%%%%%%%%%%%%%%%%%%%%%%%%%%%%%%%%%%%%%%%%%%%%%%%%%%%%%%%%%%%%%%%%%%%%%%%%%%%%%%%%%%%%%%%%%%%%%%%%%%%%%%
%f2 #&#
%
\begin{figure}[t]

\includegraphics{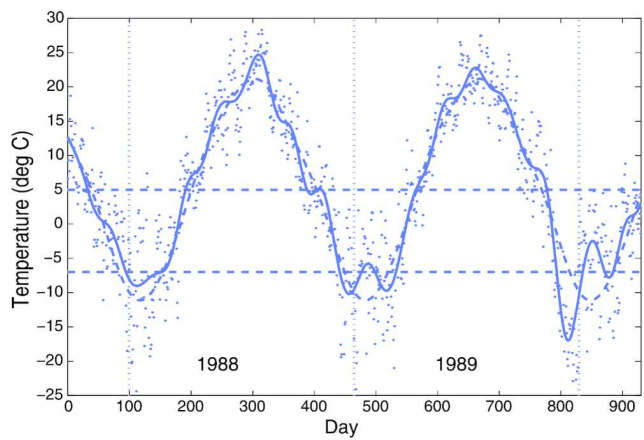}

\caption{Temperature variation in Montreal, Canada, over three
winters. The solid curve is a smooth of the daily min/max averages,
which are shown as dots. The dashed line is a strictly periodic smooth
of the data over the years 1960 to 1994. The vertical dotted lines
indicate the ``orbital'' year boundaries separated by 365.25 days. The
upper dashed horizontal line is the temperature at which growth begins
for most crops on the prairies; and the lower dashed line is the
temperature below which ice is structurally sound. Note strong
variation from year to year.}
\label{winter8889}\label{fig2}
\end{figure}
%
%%%%%%%%%%%%%%%%%%%%%%%%%%%%%%%%%%%%%%%%%%%%%%%%%%%%%%%%%%%%%%%%%%%%%%%%%%%%%%%%%%%%%%%%%%%%%%%%%%%%%%

Few time-varying events are more important than the weather.
Figure~\ref{winter8889} allows us to explore phase variation in
Montreal's daily temperature variation over three winters, winter being
the most dynamic period in the Canadian climate year. We see here
several important markers of phase variation. There are two minimum
temperatures in most winters, the first positioned around January 15
and the January thaw that separates them typically arrives on January
25. We notice, too, the increased volatility in temperature in the two
months in the dead of winter. The two horizontal lines mark
temperatures of great importance to Canada's economy. The five degree
Celsius threshold is the point at which cash crops in the Canadian
prairies germinate, and their total growth depends on, in addition to
precipitation, the total number of degrees above this threshold prior
to harvest. Minus seven degrees is the threshold below which ice has
enough structural integrity to support winter river crossings and
year-round ice dams around tailing ponds for the many mines in the
north. Global warming is altering the dates at which these thresholds
are crossed. The small plateaus in the spring and fall mark out the
arrival and departure of snow, respectively. We see that winter arrived
in both 1988 and 1989 particularly early, and with an intense cold snap
in 1989, while the 1987 winter was typical in its timing. Summer phase
variation, by contrast, seems small. Predicting phase variation is of
great importance in weather prediction, crop management, and far
northern transportation.

%%%%%%%%%%%%%%%%%%%%%%%%%%%%%%%%%%%%%%%%%%%%%%%%%%%%%%%%%%%%%%%%%%%%%%%%%%%%%%%%%%%%%%%%%%%%%%%%%%%%%%
%f3 #&#
%
\begin{figure}[t]

\includegraphics{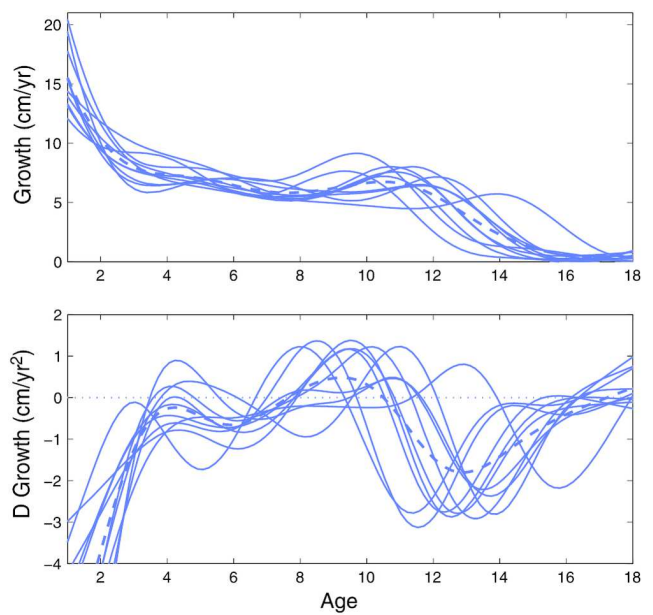}

\caption{The top panel plots the growth, understood as the first
derivative of height, of ten girls, and the bottom panel contains the
corresponding height-acceleration or growth-derivative curves. The
dashed curve in both plots is the cross-sectional mean. Both these
plots indicate both phase and
amplitude variability.}
\label{growth}
\end{figure}
%
%%%%%%%%%%%%%%%%%%%%%%%%%%%%%%%%%%%%%%%%%%%%%%%%%%%%%%%%%%%%%%%%%%%%%%%%%%%%%%%%%%%%%%%%%%%%%%%%%%%%%%

Once recognized, one sees phase variation everywhere. Parents see
children reaching puberty over a wide range of ages, and perhaps wonder
if there is some connection between the timing of the pubertal growth
spurt and adult height. Growth implies positive change, and Figure~\ref
{growth} displays the growth of ten girls in the Berkeley growth study
\citep{Tuddenham-Snyder-1954} as the positive first derivative of
height in the top panel, as well as the acceleration of height or
derivative of growth in the bottom panel. Musicians alter the timing of
notes in subtle ways to create tension and define mood, achieving in
this way their unique auditory signature as performers. Golfers and
baseball players, on the other hand, tend to find phase variation in
their swings to be an impediment to fine control over amplitude
variation, and train to the point where it is nearly eliminated.

%s1.2 #&#
\subsection{Clock Time, System Time, and the Time-Warping Function}\label{sec12}

We can articulate the concept of phase variation by distinguishing
between \emph{clock time} $s$ and \emph{system time} $t$. That is, we
envisage the spectra of wines, the weather, and children as evolving
over their respective continua at variable rates determined by
processes that we may at least partially understand and would like to
know more about. Consequently, when large-scale phase variation is
compared to the clock time, defined these days in terms of the number
of oscillations of the cesium atom, we envisage a functional
relationship $s = h(t)$ that can vary from one wine type to another,
over successive winters, and across children even within the same
family. However, the system times are defined so that all girls will
reach puberty at the same age.

In most cases, we can expect that the mapping $h$, often called the
\emph{time warping function}, will be smooth and strictly increasing,
two properties captured in the term \emph{diffeomorphism}. In other
words, we require that the inverse function value $h^{-1}(s)$ exists
everywhere in the support of the functional data since we need to use
$t = h^{-1}(s)$ to align a feature such as the pubertal growth spurt
across multiple curves. As statisticians, we look for ways to estimate
the $h$'s associated with different units of data distributed over the
base continuum, as well as ways of using discrete and continuous
covariate observations to explain and predict them.

Other conditions such as specified boundary behavior are added as makes
sense for the context at hand. For example, the time taken to produce a
sample of handwriting will vary from replication to replication, so
that $h_i$ may map, say, the interval $[0,T_0]$ into the interval
$[0,T_i]$ where $T_0$ is a fixed template time. But if the observation
is also supposed to reflect \emph{when} the handwriting event took
place, then simple shifts, $h_i(t) = t + \delta_i$, will provide a
better model. If the process under study may reasonably be expected to
have one or more derivatives, then the chain rule requires that $h$,
too, be differentiable to the same extent. In any case, it seems
unlikely that in many real-world applications
the problem constraints will allow for sharp jumps in $h$, so that
smoothness can be added to monotonicity as a property.

The following single-parameter expression for $h$ mapping $[0,T]$ into
itself serves as an illustration and is often useful:
for $\beta\neq0$,
%e1 #&#
%
\begin{eqnarray}
\label{oneparh} h(t\mid \beta) &=& T \biggl[\frac{e^{\beta t} - 1}{e^{\beta T} - 1} \biggr]\quad\mbox{and}
\nonumber\\[-7pt]\\[-7pt]\nonumber
h^{-1}(s\mid \beta) &=& \frac{1}{\beta} \log \biggl[\frac{s(e^{\beta T} -
1) + T}{T}
\biggr].
\end{eqnarray}
The expression converges to the identity warp $h(t) = t$ as $\beta
\rightarrow0$. This model, taken from \citet{Kneip-Ramsay-2008}, can also
be derived from a later
equation [equation (\ref{logD})] by setting the function $W(t) = \beta t$.

%%%%%%%%%%%%%%%%%%%%%%%%%%%%%%%%%%%%%%%%%%%%%%%%%%%%%%%%%%%%%%%%%%%%%%%%%%%%%%%%%%%%%%%%%%%%%%%%%%%%%%
%f4 #&#
%
\begin{figure}[t]

\includegraphics{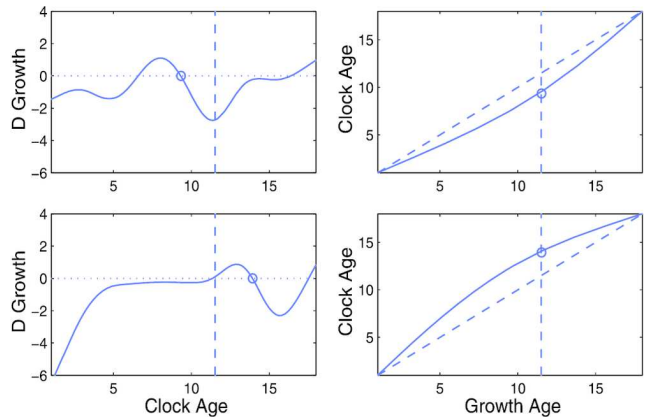}

\caption{The top left panel displays the derivative of growth for a
girl with an early growth spurt, and the bottom left panel for a girl
with a late growth spurt. The top right panel plots a warping function
$h$ that maps the growth time of the pubertal growth spurt, indicated
by the circle, into the early clock time in the left panel. The bottom
right panel shows the corresponding warping function for the late
growth spurt. This shows how phase variation is effectively modeled by
warping functions.}
\label{GrowthWarps}\label{fig4}
\end{figure}
%
%%%%%%%%%%%%%%%%%%%%%%%%%%%%%%%%%%%%%%%%%%%%%%%%%%%%%%%%%%%%%%%%%%%%%%%%%%%%%%%%%%%%%%%%%%%%%%%%%%%%%%

Some warping functions corresponding to early and late growth spurts
are shown in Figure~\ref{GrowthWarps}. The warping function
[of the type given in equation (\ref{oneparh})] in each right panel maps
pubertal growth spurt on the growth (system)
time scale into the clock or observed time scale, as indicated by the
zero crossing of the growth derivative function in the left panel, that
is, the peak of the spurt, as shown by a circle. The early pubertal
spurt in the top panel is modeled using an $h$ which moves quickly
through early growth phases relative to clock time (i.e., curves
downward) so as to produce the early clock time of about nine years,
whereas the bottom panel is better modeled with an upward curving $h$
that reflects slower transition through early growth phases to reach
the clock time of the late growth spurt of about 14 years.

%
%%------------------------------------------------------------------------------------------------

%s1.3 #&#
\subsection{The Problems that Come with Ignoring Phase Variation}\label{sec13}
\label{subsecphaseproblems}

The presence of phase variation can play havoc with classical data
analyses that are designed for data structures without phase changes.
The heavy dashed line in Figure~\ref{ethanol} is the average of the
ethanol peaks across forty wine samples, of which 31 are red. The
heights of the mean peaks are lower than almost all corresponding
sample peaks, their widths are substantially wider, and no sample peak
displays the step in the middle of the down-slope of each average peak.
That is, a statistical analysis as elementary as averaging takes the
data well outside of their normal modalities of variation, causing it
to fail as an effective data summary. A~recent review of chemometrics
\citep{Lavine-Workman-2013} highlights the importance of aligning
peaks in spectral data as a first step, and warns spectroscopists that
getting this step right can be crucial to the quality of subsequent
analyses. In fact, most familiar data analyses are found to fail in the
presence of phase variation; variances are inflated, fits by regression
models are degraded, and additional principal components are required.

This paper began as a follow-up to a workshop on curve registration at
the Mathematical Biosciences Institute at the Ohio State University in 2012
[see \citet{Marron-Ramsay-Sangalli-Srivastava-2014-EJS} and companion papers].
An effective workshop raises many more questions than it answers, and
this workshop left us with much to consider. Is there a clear
distinction between amplitude and phase variation, or is there
variation that can be represented either way? Can the transformation
$h$ be considered as a full data object, or does it just represent
nuisance variation to be discarded once identified? When phase data
objects are meaningful, how can we incorporate known covariates, such
as pH in the NMR context, into the estimation? Are ``features'' in a
curve or surface always things like peaks, points of inflection, and
threshold crossings, or can models define more general properties that
become invisible on the model side of the equation when phase is
properly incorporated and estimated? Are traditional fitting criteria
such as error sums of squares still useful, or are they only usable
when there is no phase variation? What role should derivatives play?
Can the warping function $h$ be as complex as is required to align
features, or is it wise to impose some regularity? When is it useful to
develop data analyses that reveal aspects of the \emph{joint}
variation in phase and amplitude? We will discuss some of these
questions in this paper.

Section~\ref{secgoals} defines some possible goals for curve and
surface alignment or registration, and discusses ways of understanding
what is amplitude and phase variation. Section~\ref
{secclassicalmethods} considers various optimization strategies and
statistical models that separate phase and amplitude variations.
Section~\ref{sec4} provides some links for downloading relevant softwares.
% Joint variation in phase and amplitude also raises some deeper
%mathematical issues that are taken up in Section
%\ref{secfunctionspaces}.
Section~\ref{secdiscussion} considers what has been learned in
working with these and other data sets, and looks forward to future
research and generalizations in this fascinating area.

%
%%------------------------------------------------------------------------------------------------
%%------------------------------------------------------------------------------------------------

%\newpage

%s2 #&#
\section{Viewpoints and Goals}\label{sec2}
\label{secgoals}%%

%
%%------------------------------------------------------------------------------------------------

%\begin{center}
%{\Large Outline}
%\end{center}
%
%\begin{enumerate}
%% \item object-oriented data analysis
%% \item new data objects:
%% \begin{enumerate}
%% \item warping functions for phase
%% \item aligned functions for amplitude
%% \end{enumerate}
%% \item Choices for warping functions
%% \begin{enumerate}
%% \item linear shifts
%% \item dilations
%% \item general diffeomorphisms \emph{Introduce log-derivative
%formulation here? (Jim)}
%% \end{enumerate}
%% \item Challenge of identifiability
%% \begin{enumerate}
%% \item not univocal
%% \item depends on application
%% \end{enumerate}
%% \item Interests in analysis of phase and amplitude:
%% \begin{enumerate}
%% \item amplitude is the focus; phase is a nuisance
%% \item phase is the focus; amplitude is a nuisance
%% \item joint variation is interesting
%% \end{enumerate}
%% \item Hiding phase variation by phase-plane plotting
% \item Identifiability issue
% \item Three situations:
% \begin{itemize}
% \item amplitude focus
% \item phase focus
% \item joint focus
% \end{itemize}
% \item Model dependence
% \begin{itemize}
% \item cross-sectional mean
% \item template
% \item model fit
% \end{itemize}
% \item Identification depends on what is considered known or of
%theoretical interest
%\end{enumerate}

%
%%------------------------------------------------------------------------------------------------

%\begin{center}
%{\Large Text}
%\end{center}

%
%%------------------------------------------------------------------------------------------------

%s2.1 #&#
\subsection{The Identification of Phase Variation}\label{sec21}
\label{subsecidentifiability}
In this paper we will use $y_1, y_2, \dots,$ to denote the observed
functions with both phase and amplitude variability and
$x_1, x_2, \dots,$ to be the underling functions denoting only the
amplitude variability, that is, after removing phase variability, such that
$x_i(t) = y_i(h_i(t))$.

An important challenge is \emph{identifiability} of amplitude and
phase variation, since which is which is apt to depend very much on
prior intuitions and knowledge about how each type of variation is
caused. For example, while it may seem obvious that the peaks after age
eight in the top panel of Figure~\ref{growth} exhibit phase variation,
a close look at the lower panel shows that a number of the
growth-derivative functions display more than one negative slope
episode prior to the final crossing of zero. What we are tempted to
call early spurts may only be due to the presence of a single
pre-pubertal spurt, and a late spurt may be due to two or even more
pre-pubertal spurts. This tends to sound more like an
amplitude-oriented explanation.

%f5 #&#
%
\begin{figure*}[t]

\includegraphics{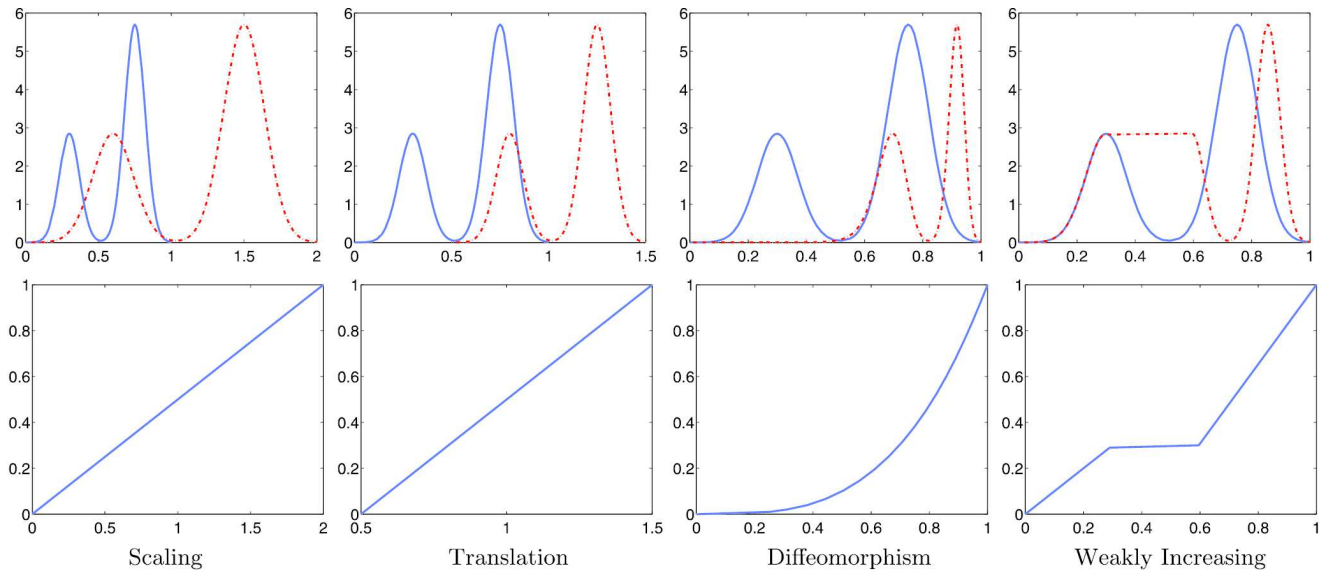}

% Scaling & Translation & Diffeomorphism & Weakly Increasing
\caption{Illustrations of different types of warping functions applied
to the same function $y$.
The top row shows $y(t)$ (solid line) and $y(h(t))$ (dashed
line), and the bottom row shows the corresponding
warping functions $h(t)$.} \label{figphase-type-examples}
\end{figure*}

A simple example of this is a data set of linear functions on $\mathbb
{R}$, $y_1, \ldots, y_n$, having the same slope, but differing
intercepts. Using the notation $x(t) = y[h(t)]$, that mode of variation
could be entirely modeled as linear shifts, $h_{i}(t)=a_{i}t+b_{i}$
constructed so that $x_{1}=x_{2}=\cdots=x_{n}$ (i.e., all variation is
in the phase variation), or it could equally well be modeled as
$h_{i}(t)=t$, the identity warp, with all of the variation in the
original data appearing in the intercepts of the $y_{i}$, or the
variation could be split between these modes.

We have tied phase variation in the wine data to a known causal factor,
the pH level of the wine, but, for the weather data, it seems to depend
on intuition as to whether spring came late in a particular year or
whether that year was simply unusually cold. Even an early velocity
peak defining the pubertal growth spurt can be seen in part as a year
of strong growth followed by a year of weaker growth. It is not
surprising, as a consequence, that we see very little attention given
to the phase variation in the evolution of statistical methodology. In
particular, the distinction between phase and amplitude variation is
generally not univocal, but instead depends on both the application
under study and the goals of a particular analysis.

%
%%------------------------------------------------------------------------------------------------

%s2.2 #&#
\subsection{Types of Phase Variations}\label{sec22}
We have mentioned the linear shifts earlier, but there are several
possibilities when choosing a class of warpings to specify phase
variation. Depending on the application context, one may prefer one
class over the others. We enumerate some possibilities below and
illustrate an example of each in Figure~\ref{figphase-type-examples}:
\begin{itemize}
\item \textit{Uniform Scaling}: Here the warping of the time domain simply
rescales it by a positive constant
$a \in\real_+$, that is, $h(t) = at$ for all $t \in\real_+$.

\item \textit{Uniform Shift}: In this case the time axis gets shifted by a
constant $c \in\real$, that is,
$h(t) = c + t$.

\item \textit{Linear or Affine Transform}: A combination of the previous
two leads to a linear or affine transformation:
$h(t) = c + at$, $a \in\real_+$ and $c \in\real$.

\item \textit{Diffeomorphisms}: A general class that includes domain
warpings is given by
the set of diffeomorphisms of the domain to itself. While it is
possible to define diffeomorphisms on the full real line, practical
considerations make it interesting to restrict warpings to compact
intervals. The set of linear transformations is contained in the set of
diffeomorphisms if the domain is defined to be the full real line.
\end{itemize}

While these are the main types of warping transformation, one can
further enlarge the scope by including functions that allow for
some flat regions; an example is shown in the rightmost column of
Figure~\ref{figphase-type-examples}. Please refer to \citet
{Srivastava-et-al-2011-arXiv}
for a discussion on the need for such functions and a rigorous approach
to handling them.

%s2.3 #&#
\subsection{Some Goals for an Amplitude/Phase Analysis}\label{sec23}
\label{subsecgoals}

We can distinguish three motivations for a model that allows for phase
variation.
First, amplitude variation could be the main focus, with phase
variation being a nuisance to be removed and then cast aside. The wine
NMR spectra in Figure~\ref{ethanol} illustrate this nicely, in part
because the goal of the analysis is specifically to model the relative
heights of the clearly visible peaks, the widths of which tend to be
proportional to their height. Prairie crop scientists tend to focus on
the total heat and precipitation available to plants in the growing
season as predictors of crop yield, leaving the issue of when the
season starts and finishes to the producers to wrestle with.
Auxologists, who study human growth, may be preoccupied by the
variation in the shape characteristics of growth curves such as the
variation in their amplitudes, and see the variation in the timings of
the pubertal growth spurt as a nuisance to be eliminated by lining up
the corresponding peaks.

On the other hand, phase variation could instead contain all of the
interesting information, in contexts where issues such as timing are
more important than relative peak heights, such as the locations of
bursts in neuronal spike train data. Crop producers know that they have
little control over heat and precipitation budgets, but they can look
for indicators of when they can sow their seeds and when certain
variants will mature. In this situation, the time warp functions are
the center of attention.

Finally, both amplitude and phase variation, and in fact the \emph
{joint variation} between these, can be central issues in the analysis.
It turns out, for example, that there is a simple relation between the
strength of a pubertal growth spurt and its timing, namely, that early
spurts are stronger and later ones are weaker, resulting in adult final
heights that do not depend much on either factor. That is, it appears
that each child has a wired-in capacity for growth, but that the
distribution of the expenditure of the growth energy over time can vary
over children with similar growth capacities.

%
%%------------------------------------------------------------------------------------------------

%s2.4 #&#
\subsection{The Role of the Model in the Amplitude/Phase Partition}\label{sec24}
\label{subsecmodel}

Assuming the relevance of phase variation, it will be clear that both
its nature and estimation strategies will depend critically on the
model being proposed for the data. The cross-sectional mean is often
the model of choice in feature alignment strategies; peaks and
threshold crossings are considered aligned when the mean curve is
centrally located within the registered curves at all points over the
interval of observation. More generally, the mean can be taken as one
of many \emph{template} or gold-standard curves to be approached as
closely as possible in some sense by the application of phase transformations.
Alternatively, as described in the next section, one can compute the
mean under a different metric and use that as a model
for alignment.
Finally, functional linear equations, low-dimensional principal
component representations, differential equations, and many other
mathematical structures may provide model spaces for amplitude
variation that, simultaneously, identify what is meant by phase
variation. That is, if a diffeomorphic transformation of the substrate
of the data, possibly within some predefined class, can improve the fit
of the model to the data, we define it as phase. Models, of course, are
usually chosen to represent a conjecture or hypothesis about what
generates the data, and in this sense the identification of the
amplitude/phase dichotomy is very much centered on the science
underlying the application.

%
%%------------------------------------------------------------------------------------------------

%s2.5 #&#
\subsection{Amplitude/Phase Separation via Equivalence Classes}\label{sec25}\label{subsecequivalence}

One way to study amplitude and phase variation is through equivalence classes.
The use of equivalence classes is not new to statistics. In fact, they
form the core idea in statistical
shape analysis \citep{mardia-dryden-book}
and in Grenander's work on pattern theory \citep
{grenander-patterns-1993}, including its applications
to computational anatomy \citep{grenander-miller98}.
In Kendall's shape analysis
the experimental units are configurations of (landmark) points in an
appropriate space, usually two- or three-dimensional Euclidean space.
To focus the analysis on the \emph{shape} variation in the data, nonshape
aspects, such as location, rotation, and perhaps scaling, are
incorporated into equivalence classes, where point configurations are
identified with each other (i.e., called equivalent) when they can be
translated, rotated, and
scaled into each other. Then, one compares shapes of objects by
comparing their
equivalence classes.
While the past shape approaches were restricted
to point sets and simple transformations (rigid motions and global
scales), the more
recent literature has studied continuous curves
with transformations that include time warpings (more precisely,
reparameterizations)
[see \citet{younes-michor-mumford-shah08} and \citet
{srivastavaetalPAMI10}, among others].

\begin{figure}[b]

\includegraphics{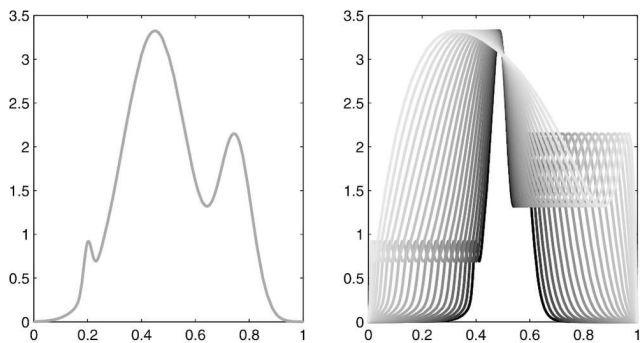}

\caption{Different time warps of a function (left) form an equivalence
class from the
perspective of defining its amplitude.} \label{figequivalence-illustrate}
\end{figure}

In an entirely parallel fashion, one can define amplitude and
phase variability in functional data using equivalence classes. As laid
out in \citet{Srivastava-et-al-2011-arXiv},
the main idea is to understand
amplitude variation through a quantity that incorporates
all aspects of phase variation inside it. This is done by defining
an equivalence relation, where curves are identified or deemed
equivalent when they can be time warped
into each other. Figure~\ref{figequivalence-illustrate} shows
some elements of an equivalence class---a set of warps of a single
three-peak curve. The equivalence class
is actually much bigger, including all diffeomorphic time warps of
this curve, only some of which are shown here. These equivalence
classes are now taken as representing \emph{amplitudes} because
they model the essence of vertical variation in a simple and natural
way. The phase variation is incorporated \emph{within equivalence
classes}, while the amplitude variation appears \emph{across equivalence
classes}. Further motivation for how equivalence classes provide clear
definitions for
separation of amplitude and phase is given in Section~\ref{sec34}.
(See also \cite{Vantini2012}.)

%f6 #&#
%

While the origins of these ideas lie in shape theory,
an understanding of these concepts can also be obtained using the
terminology of object-oriented data analysis (OODA),
as defined in \citet{Wang-Marron-2007} and more recently discussed in
\citet{Marron-Alonso-2014}.
An important special case of OODA is FDA, where functions are the data
objects. A natural approach to the decomposition of amplitude and phase
variation is to model each with appropriate data objects, with specific
goals as laid out in Section~\ref{sec23}.
In some situations, such as the wine NMR data in Figure~\ref{fig1}, the phase
variation can be viewed as a nuisance, so the data objects of interest
are registered curves, that is, time warped to match their peaks. In
other situations, for example, the temperature data shown in Figure~\ref{fig2}
and for human growth curve data in Figure~\ref{fig4},
interesting data objects can be any of the registered amplitude curves,
or the transformations used to achieve registration (reflecting phase
variation), or else the concatenation of both, for situations where
joint amplitude--phase variation is key.
In the same spirit, the data objects in an equivalence-class approach
are the equivalence classes themselves.

%s3 #&#
\section{Some Current Curve Registration Methods}\label{sec3}\label{secclassicalmethods}

In this section we look at a few curve registration techniques for
estimating warping functions $h$. In the first two sections, the focus
is on using a template function $x_0$ as a target, so that $y(s)
\approx x_0[h(t)]$ and, inversely, $x_0(t) \approx y[h^{-1}(s)]$. We
will see that the sense in which the approximation is defined requires
considerable care, with least squares approximations
computed in the usual way not being a viable candidate. The template
$x_0$ is often defined using an objective function
whose solution is iterative, starting with the cross-sectional mean and
alternating between a registration and a recalculation of the
cross-sectional mean
of the registered functions. Typically this process, often referred to
as \emph{Procrustes iterations}, converges in only a few steps.

%
%%------------------------------------------------------------------------------------------------

%s3.1 #&#
\subsection{Dynamic Time Warping (DTW)}\label{sec31}
\label{subsecDTW}

Before examining current registration methods, it is worthwhile
mentioning dynamic time warping (DTW), an early
registration method applied to discrete sequences of phonemes (a basic
unit of language). \citet{Sakoe-Chiba-1978} devised an
insertion/deletion algorithm that is rather like that of isotonic
regression \citep{barlow1972statistical}. The underlying
algorithm, which is a dynamic programming algorithm, is an optimization
technique where one
partitions the graph space using a finite grid and the warping $h$ is
restricted to be a piecewise-linear function passing through the
nodes of this grid. Depending on the context, one may allow it to have
vertical jumps or be horizontal for multiple time-steps.
In the classical DTW, the dynamic programming algorithm is applied to
minimize the least-squares cost function
given in equation (\ref{eqleast-squarle}).
DTW can be effective as a feature alignment method, as it provides a
globally optimal solution, albeit on
the restricted search space (piecewise-linear $h$ on a fixed grid). But
the classical DTW has the conceptual problem that it may not provide
smooth differentiable time warps that many applications require.
Also, the computational algorithm can be greedy, in the sense of
warping regions where no alignment seems called for.
These problems, in general, can be handled by adding a regularization
term to the cost function.

%
%%------------------------------------------------------------------------------------------------

%s3.2 #&#
\subsection{Landmark Registration}\label{sec32}
\label{subseclandmark}
In terms of functional data alignment,
we begin with the easiest situation in which each curve $y_i(s)$ has
clearly-defined features, the timings of which can be used to estimate
$h_i$ at a series of points $(t_\ell, h_{i \ell})$. This requires, in
turn, a consideration of what we might mean by ``feature.''

In the case of the wine data, there seems to be little confusion. In
most types of spectra, the presence of a chemical compound is marked by
a single peak, the location of which is the desired landmark, and
automatic methods for peak detection are relatively easy to devise. For
multi-peak structures such as the NMR peaks in Figure~\ref{ethanol},
the average of the peak locations would serve the purpose.
Alternatively, a template can be set up for a peak shape, and a peak
detector can be devised by computing correlations with moving windows
of the curve shape with the template pattern.

Let us suppose that there is a gold standard template spectrum $x_0$
with $L$ peaks occurring at times $t_\ell, \ell= 0,\ldots,L+1$,
where times $t_0$ and $t_{L+1}$ are the endpoints of the observation
interval. Then, for the $i$th spectrum with peak locations at $s_\ell
$, we can estimate $h_i$ by interpolating in some suitable way the
pairs $(s_\ell, t_\ell)$. Polygonal lines might serve, or it may be
important to use a smoother interpolant having, perhaps, a specified
number of derivatives. Figure~\ref{GrowthWarps} offers an elementary
example of landmark registration, where the timing of a girl's pubertal
growth spurt is the single landmark $t_i$,
shown as a circle in Figure~\ref{GrowthWarps}, and the intervals $(3,
t_i)$ and $(t_i, 18)$ for the $i$th girl are interpolated by the
warping functions
[formed using the expression in equation~(\ref{oneparh})].

Peak and valley locations can be translated into crossings of zero in
the curve's first derivative with negative and positive slopes,
respectively. Other types of crossings may also be important. For
example, the heavy-duty winter in Figure~\ref{winter8889} can be
defined as the average of the first crossing time with negative slope
for $-$7 deg C and the second crossing time with positive slope. Prairie
farmers would prefer the crossing of germination threshold of 5 deg C
with positive slope, and, in fact, do just that with daily soil
temperature readings in May.

The problem with landmarks, of course, is that they are not always
visible or one may be faced with other types of
feature time ambiguity such as two or more closely spaced $-$7 deg C
crossings in the temperature data. Moreover, recording landmarks by
hand is tedious, and fail-safe automatic detectors are sometimes hard
to set up. The choice of landmark can itself be open to the kind of
debate that scientists would prefer to avoid. Finally, landmark
registration is only discrete evidence concerning the intrinsically
continuous function $h_i$, and as such ignores what happens in between
landmarks, where there may reside additional information about $h$.

%
%%------------------------------------------------------------------------------------------------

%s3.3 #&#
\subsection{Registration Using $\mathbb{L}^2$ Distance and
Correlational Criteria}\label{sec33}
\label{subsecL2critera}

Now we look at a classical approach to functional registration that
does not require the use of landmarks.
Let $h_i$ denote the time warping associated
with the $i$th data item $y_i$;
this $h_i$ can be restricted to be an element of a parametric family,
defined by the value of one or more parameters, or can be fully
nonparametric as in a diffeomorphism.
The one-parameter warps [equation~(\ref{oneparh})], along with simple
shifts, scale changes, and linear functions of $t$, are examples of
simple parametric warping families, and we will propose more flexible
representations in Section~\ref{sec4}. It seems natural to specify a loss
function $L$ that optimizes the congruence of a set of clock-time
functions $y_i$ to corresponding warped versions of a template $x_0$,
that is, $y_i\approx x_0 \circ h_i$, where $(x_0 \circ h_i)(t)=x_0(h_i(t))$.

\begin{figure}[t]

\includegraphics{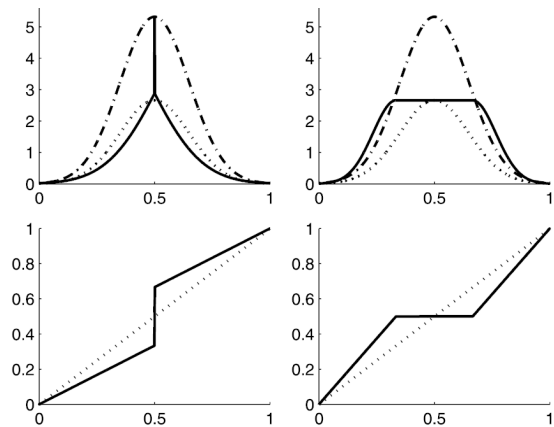}

\caption{The upper panels show a Gaussian density function $x_0$ and
its scaled version $y$, as dot-dashed and
dotted curves, respectively. The solid curve in the upper left panel
results from minimizing the squared error criterion $\int[y(t) - (x_0
\circ h)(t)]^2 \,dt$
with the optimal warping function $h$ shown in the lower left panel.
The solid curve in the upper right panel results from minimizing the
squared error criterion $\int[(y \circ h)(s) - x_0(s)]^2 \,ds$ with the
optimal warping function $h$ shown in the lower right panel.}
\label{L2Warp}
\end{figure}

The choice, however, of standard options such as
%e2 #&#
%
\begin{eqnarray}\label{eqleast-squarle}
L(h; y_i, x_0) &=& \llVert
y_i - x_0 \circ h_i \rrVert ^2
\nonumber\\[-8pt]\\[-8pt]\nonumber
&=& \int\bigl[y_i(t) - x_0\bigl( h_i(t)
\bigr)\bigr]^2 \,dt
\end{eqnarray}
will quickly prove disappointing if combined with a flexible class of
warping functions, as Figure~\ref{L2Warp} demonstrates.
In the left case, the minimization of the $\mathbb{L}^2$ norm results
in a
reduction from 0.500 to 0.024, using a
piecewise-linear warping and a spike that nearly eliminates the area
under the registered curve corresponding to intervals
where the $y$ has larger amplitude than $x_0$. In the registration
process the amplitude characteristics of $y$ have been significantly distorted.

%f7 #&#
%

%\emph{Comment: The warps shown in the figure are not optimal in the L2
%sense given an arbitrarily flexible warping family. The warp shown in
%the earlier version was optimal everywhere except in a small
%neighborhood of 0.5 due to numerical issues, and this could have been
%fixed post-hoc in the Matlab code to give the sharp spike in the upper
%left. Likewise with the upper and lower right figures. I'll produce
%these figures for your consideration later. JOR}

This pinching effect can be mitigated by using warping functions that
are constrained to be smooth, either by the use of a regularization
strategy or by the use of a small number of basis functions. The
registration procedures proposed in \citet{Ramsay-Silverman-2005}, for
instance, incorporate a penalization term that forces the choice of the
warping functions toward functions that do not differ significantly
from the identity (corresponding to the case of no registration) or
from constant functions. Concerning instead the use of simple
parametric families for the class of warping functions, the $\mathbb{L}^2$
distance will work just fine for the one-parameter shift-warp family,
$h(t) = t + \delta$. Such a
registration procedure performs perfectly for the example in
Figure~\ref{L2Warp}, where the identity warp is returned since the two peaked
curves are already registered.

It thus appears fundamental to appropriately relate the definitions of
amplitude variation and of phase variation, that are jointly described
by the loss function to be optimized and the class of warping functions.
%The concepts of amplitude and phase variation cannot be modeled
%independently
%of one another. On the contrary, they need to be defined
%simultaneously to avoid issues such as the one highlighted
This motivates the simultaneous definition of phase and amplitude to
avoid issues such as the one highlighted
in Figure~\ref{L2Warp}. For instance, the loss function $L$ to be
optimized and the class of warping functions $h$ may be chosen so that
for any two functions $x_1$, $x_2$, and any warping function $h$, $L$
satisfies the relation
%e3 #&#
%
\begin{equation}
\label{H-invariance} L (x_1, x_2 ) = L (x_1 \circ
h, x_2 \circ h ).
\end{equation}
This invariance property guarantees that it is not possible to obtain a
fictitious increment of the similarity between two functional data by
simply warping them simultaneously with the same warping function, and
has been clarified in the context of different types of warpings in
different papers.
For example, \citeauthor{Sangalli-Secchi-Vantini-Veneziani-2009-JASA} (\citeyear{Sangalli-Secchi-Vantini-Veneziani-2009-JASA,Sangalli-Secchi-Vantini-Vitelli-2010-CSDA})
and
\citet{Vantini2012} study this invariance and then specify it in the
context of linear or affine transformations of the domain, while \citet
{Srivastava-et-al-2011-arXiv} study it for diffeomorphisms.

%f8 #&#
%
\begin{figure*}[b]

\includegraphics{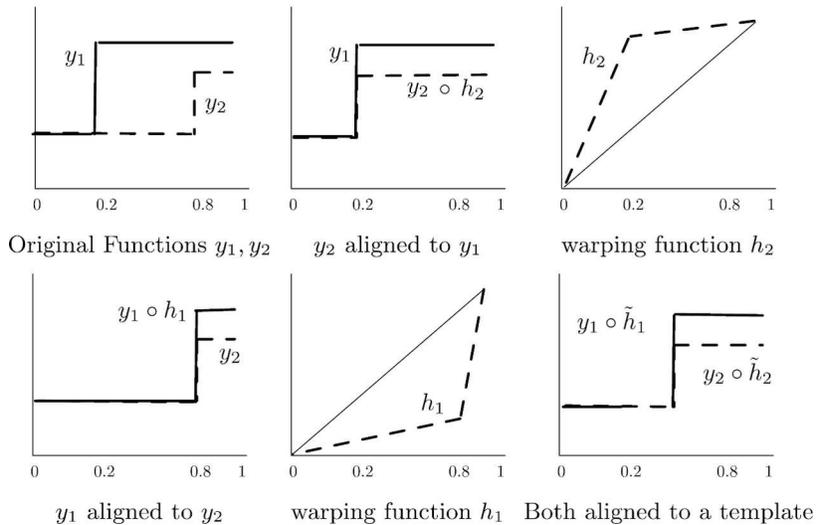}

%Original Functions $y_1, y_2$ & $y_2$ aligned to $y_1$ & warping
%function $h_2$\\
%$y_1$ aligned to $y_2$ & warping function $h_1$ & Both aligned to a
%template\\
\caption{Toy example showing asymmetry of the $\mathbb{L}^2$ norm naively
applied to curve registration.} \label{figSteveFigureB}
\end{figure*}

Moreover, as already highlighted, the concepts of amplitude variation
and of phase variation are problem-specific and depend on the
application goals. For instance, if two functional data $x_1$ and $x_2$
may be considered aligned when they are proportional, that is, when
$x_1=\alpha x_2$, then it is natural to use the loss function
associated to the semi-norm
%e4 #&#
%
\begin{equation}
\label{eqL2-normalized} \biggl\llVert \frac{x_1}{\llVert  x_1\rrVert   } - \frac{x_2}{\llVert  x_2\rrVert   } \biggr\rrVert
,
\end{equation}
and the corresponding correlation measure
%e5 #&#
%
\begin{equation}
\label{rhocrit} \rho(x_1,x_2) = \frac{ \langle x_1,x_2  \rangle}{\sqrt
{ \langle x_1,x_2  \rangle \langle x_2,x_2
\rangle} }.
\end{equation}
The class of linear warping functions $h(t)=\delta+\gamma t$ is
compatible, in the sense of equation (\ref{H-invariance}), with the loss
function associated to (\ref{eqL2-normalized}) and (\ref{rhocrit}),
if these are computed over the full real line. This definition of
amplitude/phase variation seems, for instance, well suited for the wine
data, where the amplitude variation is well described by the relative
heights of the peaks, rather than by their absolute heights, and where
linear transformations of the abscissa allows for a good alignment of
these peaks.
This also holds for growth curve data, where the emphasis is on growth
velocities, rather than on the height curves per se, and the children's
biological clocks, with their pubertal spurts, are aligned by aiming at
proportional growth velocities.
\citet{Ramsay-Silverman-2005} used the size of the minimum eigenvalue
of the order two cross-product matrix
%e6 #&#
%
\begin{eqnarray}\label{regcriterion}
&& L(h; y, x_0)
\nonumber\\[1pt]\\[-17pt]\nonumber
&&\quad  = \lleft[ \matrix{ \displaystyle\int\bigl
\{x_0(t)\bigr\}^2 \,dt & \displaystyle\int x_0(t) y
\bigl[h(t)\bigr] \,dt
\cr
\displaystyle\int x_0(t) y\bigl[h(t)\bigr] \,dt & \displaystyle\int\bigl
\{y\bigl[h(t)\bigr]\bigr\}^2 \,dt} \rright]. %
% \left[
% \begin{array}{ll}
% \int x_0(t)^2   \,dt & \int x_0(t) y_i(h^{-1}(s\right\vert \thetabold))   \,dt \\
% \int x_0(t) y_i(h^{-1}(s \right\vert  \thetabold))   \,dt & \int y_i(h^{-1}(s\left\vert
%\thetabold))^2   \,dt
% \end{array}
% \right].
\end{eqnarray}
The minimum eigenvalue criterion essentially measures the linearity of
the relationship between $x_0$ and $y \circ h^{-1}$, and is the same
thing as maximizing the correlation [equation (\ref{rhocrit})] between
the two functions or, correspondingly, minimizing [equation (\ref{eqL2-normalized})].
In other contexts, two functional data $x_1$ and $x_2$ may be
considered aligned when their first derivatives are proportional, that
is, $Dx_1=
\alpha D x_2$ and, equivalently, $x_1=\alpha x_2+\beta$. Then it is
natural to use the loss function in equations (\ref{eqL2-normalized})
and (\ref{rhocrit}), but applied to the first derivative instead.
Also, in this case, if the loss function is computed over the full real
line, then it is compatible in the sense of equation (\ref
{H-invariance}) with the class of linear warping functions $h(t)=\delta
+\gamma t$. And the same can of course be said for the $\mathbb{L}^2$
distance (\ref{eqleast-squarle}) with the shift-warp family
$h(t)=\delta+ t$.
\citet{Sangalli-Secchi-Vantini-2014-EJS-AneuRisk-analysis} report other
examples of loss-functions/class of warping functions, that define
concepts of amplitude and phase variations that are appropriate in
different applications. In practice, the functional data are only
available on bounded intervals, that possibly differ from curve to
curve. These loss functions can then be computed over the intersection
of the domains of the two functional data. In the case of the $\mathbb{L}^2$
distance, normalizing the distance by the length of the domain
intersection helps avoiding fictitious decrements of the distance as
the intersection becomes smaller.

It is also possible to consider much more flexible representations of
phase variation and still define loss functions and class of warping
functions satisfying the property (\ref{H-invariance}). Section~\ref{sec34} is
devoted to the case where the phase variation is described by arbitrary
diffeomorphic transformations.

%s3.4 #&#
\subsection{The Square-Root-Velocity Function and the Fisher--Rao Metric}\label{sec34}
\label{subsecFRmetric}

Standard fitting criteria such as least squares may also be applied to
transformations of the functional objects, most commonly
first and second derivatives
or their combinations. However, one can go even further by choosing
newer metrics that are compatible with
the notion of equivalence classes mentioned earlier in Section~\ref{sec24}.
Application of the concept of equivalence classes as data objects
in FDA needs some rethinking of important concepts. First off, the
classical notion of metrics on curves needs to be extended to metrics
on equivalence classes. Some consideration of this point highlights
the challenges faced by classical approaches in analyzing vertical
and horizontal curve variation. For example, as mentioned in the
previous section, a common approach to
quantifying the \emph{vertical distance} between curves $y_{1}$ and
$y_{2}$ is through $\mathbb{L}^2$ norm between $y_{1}$ and warped $y_{2}$,
that is,
$\inf_{h}\llVert  y_{1}-y_{2} \circ h \rrVert _{2}$.
From a theoretical perspective this quantity has several problems: it
is not
symmetric and does not satisfy the triangle inequality.
Moreover, from a conceptual perspective, there are problems with this
formulation, as shown in Figure~\ref{figSteveFigureB} [constructed by
\citet{lu2013principal}].
The top left panel of Figure~\ref{figSteveFigureB} shows a toy
example, using two single step functions as $y_1$ and
$y_2$. One naive approach
to aligning these curves is to register $y_{2}$ to $y_{1}$ using
the simple piecewise-linear warp $h_2$ shown in the top right panel.
The result of this is a reasonable alignment shown in the
top center panel. But an equally good approach to aligning these
curves is to warp $y_{1}$ into $y_{2}$, using the alternate
piecewise-linear warp $h_1$ shown in the bottom center panel. As shown
in the bottom
left, this also gives a high quality of alignment. The challenge
in classical approaches is what should be taken as the vertical distance
between $y_{2}$ between~$y_{1}$? The (appropriately squared, etc.)
region between the aligned curves (representing the $\mathbb{L}^2$
norm) in the
top center panel
is clearly very different from that in the bottom left panel.
Now if we allow warping of both $y_{1}$ and $y_{2}$, then many other
appealing registrations could be found, for example, that in the bottom right
panel, all of which are quite reasonable. A big payoff of the idea
of equivalence classes as data objects is that it allows a very simple
and natural metric, which essentially includes all of these reasonable
alignments in its formulation.

The core idea is to choose a metric that helps compare equivalence
classes, and not just individual functions, since these classes provide
an identifiable representation of amplitude variability in
this setting. This is done in a straightforward way, by starting with a
curve metric that is invariant to identical warping of its two
arguments, as in equation (\ref{H-invariance}). That is, it should satisfy
%e7 #&#
%
\begin{equation}
d(x_1, x_2) = d(x_1 \circ h, x_2
\circ h), \label{eqFR-isometry}
\end{equation}
for all warpings $h$. This is a particularization of equation (\ref
{H-invariance}) where a general
loss function is replaced by a distance function. \citet
{Srivastava-et-al-2011-arXiv}
used the nonparametric form of the Fisher--Rao metric [see
\citet{srivastava-etal-Fisher-Rao-CVPR2007} for a short introduction to
this metric] for this purpose.
In fact, since the original Fisher--Rao metric was defined only for
positive probability densities, they extended this notion to include
a larger class of functions. The actual expression for this metric is
complicated and thus is not discussed in detail here, except we note
that the resulting Fisher--Rao distance, denoted by $d_{\mathrm{FR}}$, satisfies
the property stated in equation (\ref{eqFR-isometry}).

\begin{figure}[b]

\includegraphics{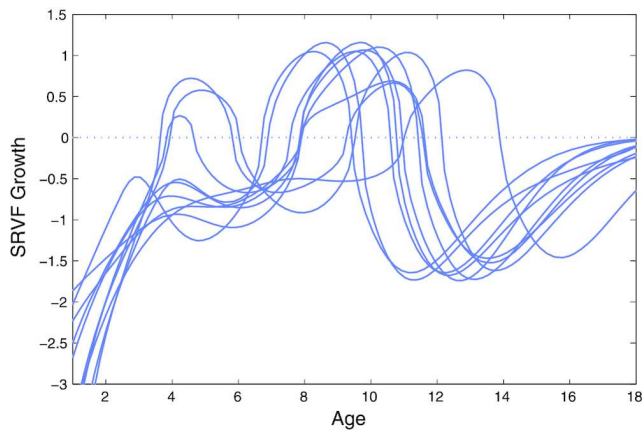}

\caption{The signed root velocity transforms of the ten female growth
curves displayed in Figure~\protect\ref{growth}.}
\label{SRVFgrowth}
\end{figure}

The key step in this formulation is to define a \emph{square root
velocity function} (SRVF) transform,
%e8 #&#
%
\begin{equation}
\label{SRVFtransform} \SRVF(x) = \sign(Dx) \sqrt{\bigl(\llvert Dx\rrvert \bigr)},
\end{equation}
where $\sign(u) = +1$ if $u \geq0$ and $-1$ if $u < 0$ and $D x$ is
the first derivative of $x$.
It should be noted that SRVF is a one-to-one map up to a translation.
That is, if $x(0)$ is known, then one
can calculate $x$ back uniquely from its SRVF.
The SRVF transforms of the ten growth curves in Figure~\ref{growth}
are shown in Figure~\ref{SRVFgrowth}. In this particular case,
since the $x$ is defined to be the derivative of growth,
$\SRVF(x)$ refers to the acceleration curves shown in the bottom panel
of Figure~\ref{growth}. Consequently, the SRVF curves cross the zero
axis at the same locations, but now with very steep slope.

%Moreover, the SRVF transform removes a great deal of the amplitude
%variation in the acceleration curves. In short, the SRVF transform is
%designed to enhance phase variation at the expense of amplitude
%variation. This seems ambiguous -- if used this need a mathematical
%justification!

%%%%%%%%%%%%%%%%%%%%%%%%%%%%%%%%%%%%%%%%%%%%%%%%%%%%%%%%%%%%%%%%%%%%%%%%%%%%%%%%%%%%%%%%%%%%%%%%%%%%%%
%f9 #&#
%
%
%%%%%%%%%%%%%%%%%%%%%%%%%%%%%%%%%%%%%%%%%%%%%%%%%%%%%%%%%%%%%%%%%%%%%%%%%%%%%%%%%%%%%%%%%%%%%%%%%%%%%%

The main reason for introducing SRVF is that the Fisher--Rao distance
between any two functions is
given by the $\mathbb{L}^2$ distance between their SRVFs, that is,
%e9 #&#
%
\begin{equation}
\label{FRmetric} d_{\mathrm{FR}}(x_1, x_2) = \bigl\llVert
\SRVF(x_1) - \SRVF(x_2)\bigr\rrVert.
\end{equation}
We refer the reader to \citet{Srivastava-et-al-2011-arXiv} for the
details, but mention in passing that
the proof hinges on the fact that $\SRVF(x \circ h) = (q \circ h) \sqrt
{ Dh}$, where $q = \SRVF(x)$.

This nice mathematical structure leads to {\em formal} definitions of
amplitude and phase in functional data.
For any two functions, $x_1$ and $x_2$, the actual registration problem
is given by
%e10 #&#
%
\begin{eqnarray}\label{eqFR-register}
&& \inf_{h} \bigl\llVert \SRVF(x_1) -
\SRVF(x_2 \circ h)\bigr\rrVert
\nonumber\\[-8pt]\\[-8pt]\nonumber
&&\quad= \inf_{h} \bigl
\llVert \SRVF(x_1 \circ h) - \SRVF(x_2)\bigr\rrVert.
\end{eqnarray}
This formulation avoids the issue discussed in the example associated
with Figure~\ref{figSteveFigureB}.
It is important to note
that $\SRVF(x \circ h) \neq(\SRVF(x) \circ h)$ and, therefore, this
alignment is NOT simply a least-square
alignment of SRVFs.
The infimum value in equation (\ref{eqFR-register}) represents a
comparison of the amplitudes of $x_1$ and $x_2$ and is actually
a distance between the equivalence classes discussed in Section~\ref{sec25}.
If the optimal $h$ on the left-hand side is invertible, then its inverse is
also the optimal for the right-hand side of that equation.
This has been called {\em inverse consistency} in the image-processing
literature. The optimal $h$ denotes the
(relative) phase between $x_1$ and $x_2$. The actual optimization over
$h$ in equation (\ref{eqFR-register}) can be performed in many ways,
depending on the problem. If $h$ takes a nonparametric form, a~diffeomorphism of the domain, then the
dynamic programming algorithm mentioned earlier is applicable. If some
application calls for smooth phases,
then some common smoothing idea---either restrict to a parametric
family or apply a regularization penalty---can be applied, both at a loss of some mathematical structure. We
emphasize that while some applications favor smooth
solutions for warpings, some others, such as activity recognition in
computer vision,
naturally favor warping functions that are close to being vertical or
horizontal over
subdomains.

%
%%------------------------------------------------------------------------------------------------
%%------------------------------------------------------------------------------------------------
%%------------------------------------------------------------------------------------------------

%s3.5 #&#
\subsection{Representations of the Warping Function}\label{sec35}
\label{subsecwarprep}
The nature of warping functions leads to some interesting representations.
The evolution of time, whether clock or system, is fundamentally a
growth process, and as such, like height, has a positive first
derivative. Two transformations of $h$ play a number of useful roles in
the representation and study of phase variation. Using the notation
$Dh$ for the derivative of $h$, the \emph{log--derivative
transformation} and its inverse
%e11 #&#
%
\begin{eqnarray}\label{logD}
&\displaystyle h(t)
=  C_0 + C_1 \int _0^t \exp W(v) \,dv,\quad C_1 > 0\quad\mbox{and} \hspace*{-3pt}&
\nonumber\\[-8pt]\\[-8pt]\nonumber
&\displaystyle (\log D) h(t) - \log C_1 = W(t)&
\end{eqnarray}
allow us to represent any diffeomorphism $h$ in terms of the
unconstrained log--derivative function $W$. A natural and effective
method of computing $h$ is to use numerical differential equation
solver methods to approximate the solution of the linear forced
differential equation $Ds = \exp[W(t)]$ using the initial value $h_0 =
C_0$. Moreover, from the equation $h^{-1}[h(t)] = t$ the solution of
the complementary nonlinear unforced equation $Dt = \exp[-W(t)]$
defines the inverse of the warping function.

Since the log--derivative $W$ is unconstrained and defined over a closed
interval, it is natural to use a basis function expansion, with the
B-spline basis being the likely choice. In particular, the
one-parameter model [equation (\ref{oneparh})] corresponds to $W(t) =
\beta t$. The overall smoothness of $h$ can be controlled either by the
number of basis functions used or by appending a roughness penalty to a
fitting criterion.
It is essential that any representation be expandable to include
contributions from one or more covariates $z_j$ known or conjectured to
modulate phase. For example, it is well known in climate modeling that
proximity to oceans retards the seasons by two to three weeks, so that
a model for phase variation across weather stations would include this
factor. Because of global warming, long-term time itself is an
important modifier of climate variables such as seasonal temperature
and precipitation. Covariates can be
easily incorporated by extending $W$ to be a function of a covariate
such as $W(t + \alpha z)$ for $W(t,z)$.

%f10 #&#
%
\begin{figure*}[t]

\includegraphics{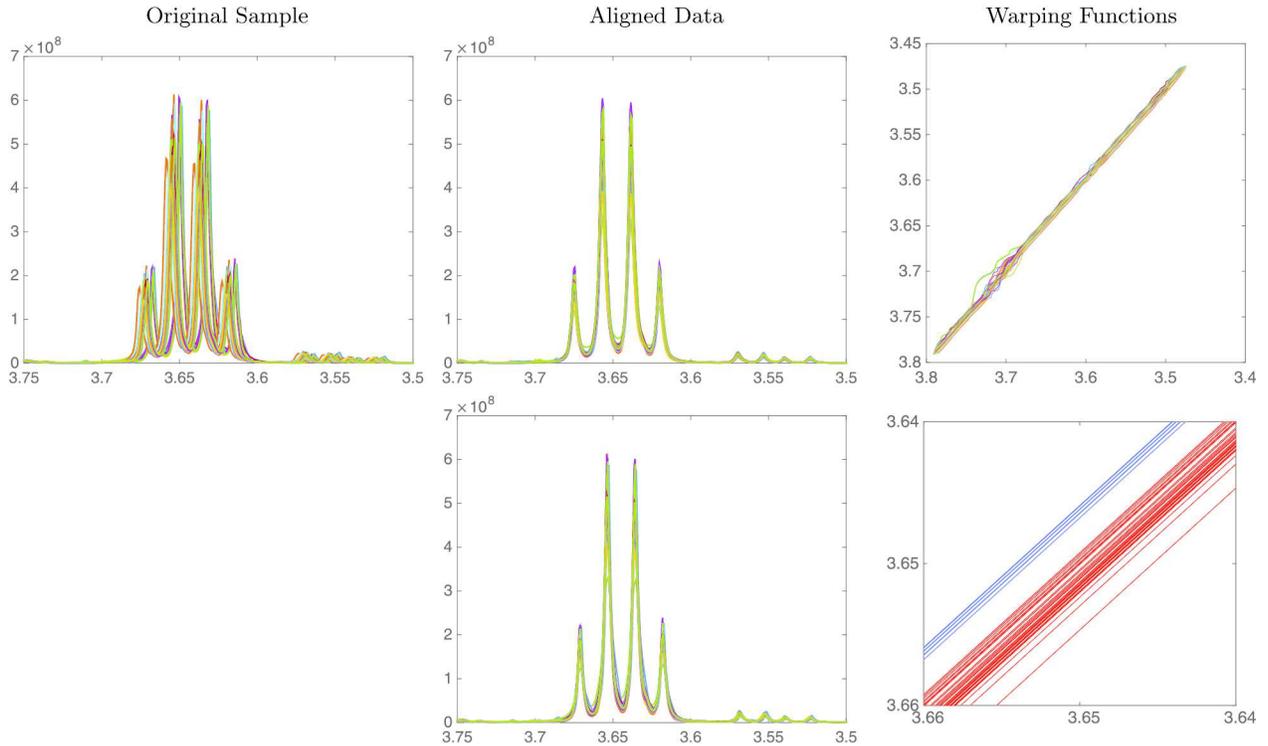}

%Original Sample & Aligned Data & Warping Functions \vspace*{0.1in}\\
\caption{Alignment of a part of 40 wine NMR spectra shown earlier.
Top row: using the SRVF framework. Bottom row: using shifts and
minimizing the loss in (\protect\ref{eqL2-normalized}). A zoom of
the warping functions, displayed on the bottom right panel, shows a
neat separation in the phase between the red wines (warping functions
colored in red) and the white and ros\'{e} wines (warping functions
colored in blue).}
\label{figwine-data-alignment}
\end{figure*}

Another mathematical representation for warping functions comes from
the SRVF idea.
Since $Dh$ is assumed to be positive, one can also use the positive
square root $\psi(t) = \sqrt{Dh(t)}$
as a representation of $h$. Just like $W$ earlier, one can use a basis
expansion to express $\psi$
if $h$ is not constrained any further. However, if $h$ represents a
time warping of a fixed interval,
for instance, $[0,1]$, to itself, then that imposes an additional
constraint on $h$. In order to
obtain the boundary\vspace*{1pt} conditions $h(0) = 0$ and $h(1)= 1$, we require
that $\int_0^1 \psi(t)^2 \,dt = 1$
or the $\mathbb{L}^2$ norm of $\psi$ is one.
This is an interesting geometric structure---the space of allowed
$\psi$ functions is a unit sphere and its geometry
can be exploited in the ensuing analysis. The spherical geometry of
this space of $\psi$ functions has been used to
perform estimation and alignment of curves in several places,
including \citet{ashok-srivastava-etal-TIP09}.
This geometry has also been helpful
in developing PCA of warping functions [\citet{Tucker-CSDA2013}] and in
alternatives to PCA in the form of principal nested spheres [\citet
{jung-etalBiometrika2012}].

%
%%------------------------------------------------------------------------------------------------
%%------------------------------------------------------------------------------------------------
%%------------------------------------------------------------------------------------------------

%s3.6 #&#
\subsection{Registering Curves to Models}\label{sec36}
\label{subsecmodelregistration}

So far we have focused on pairwise registration of functions, but the
alignment of multiple
functions is often more of concern in analyzing real data. While some
methods for multiple alignment are simple
extensions of the binary case, the others take a completely fresh
approach and derive models tailored
to such function data objects. The former approach is generally
based on constructing a \textit{template} of some kind and then
registering individual
functions to this \mbox{template}. This template may be constructed in an
iterative fashion, as
\mbox{recursive} improvements in alignments
improve the resulting template, and vice-versa.

A simple idea for constructing a template is the cross-sectional mean,
as mentioned earlier. At each iteration,
one takes the currently aligned functions $\{ x_i \circ h_i\}$ and
computes their cross-sectional mean to update the template
$x_0 = {1\over n} \sum x_i \circ h_i$. (The cross-sectional mean is, of
course, the mean of functional objects
under the $\mathbb{L}^2$ metric.)
Then, one by one, the given functions are aligned to this template to
update $h_i$'s: $h_i = \argmin_h L(h; x_i, x_0)$.
Depending on the nature of data, the results of this process may be
sensitive to the initial conditions.

The same idea can be generalized to situations where a metric different
from the $\mathbb{L}^2$ metric is used.
In the case where equivalence classes of functions are data objects,
one can
compute the average of the corresponding equivalence classes
$ [x_{1} ],\ldots, [x_{n} ]$, using the notion
of a Karcher or Fr\'echet mean. This can be done
under the Fisher--Rao distance mentioned in the previous section. The template
is then taken to be the \emph{center} of the Karcher mean equivalence class,
chosen so that the average of the phases of $x_{1},\ldots,x_{n}$, with
respect to this center,
is the identity $h_{id}$. For further details of this construction
and an algorithm for computing the center of an orbit, please refer
to \citet{Srivastava-et-al-2011-arXiv}.

%f11 #&#
%
\begin{figure*}

\includegraphics{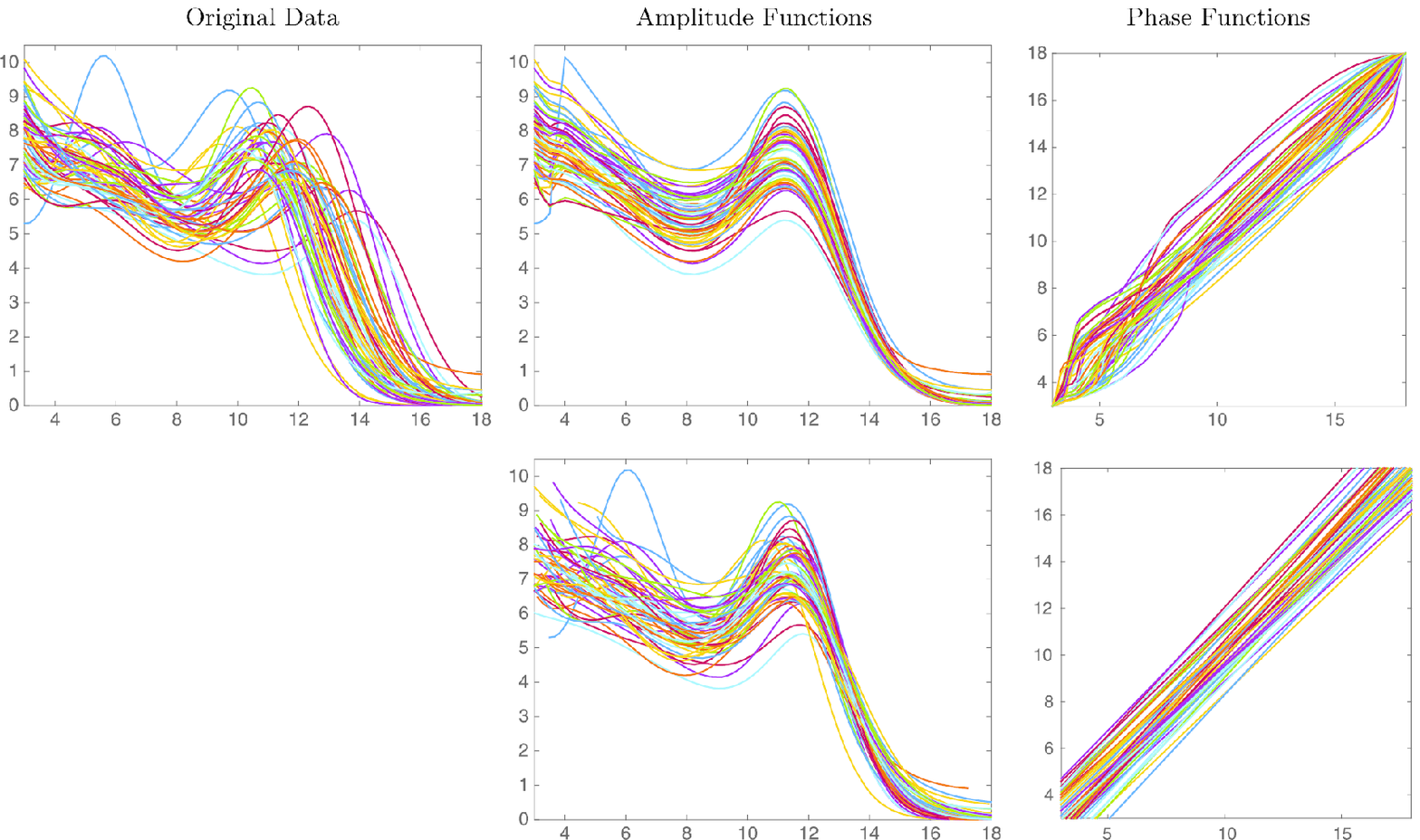}

%Original Data & Amplitude Functions & Phase Functions \vspace*{0.08in}
%\\
\caption{Alignment of the growth velocities of the 54 girls in
the Berkeley growth study using the SRVF framework (top) with the
original data and the linear
alignment (bottom).}
\label{figgrowth-female-data-alignment}
\end{figure*}

%%%%%%%%%%%%%%%%%%%%%%%%%%%%%%%%%%%%%%%%%%%%%%%%%%%%%%%%%%%%%%%%%%%%%%%%%%%%%%%%%%%%%%%%%%%%%%%%%%%%%%
%f12 #&#
%
\begin{figure*}

\includegraphics{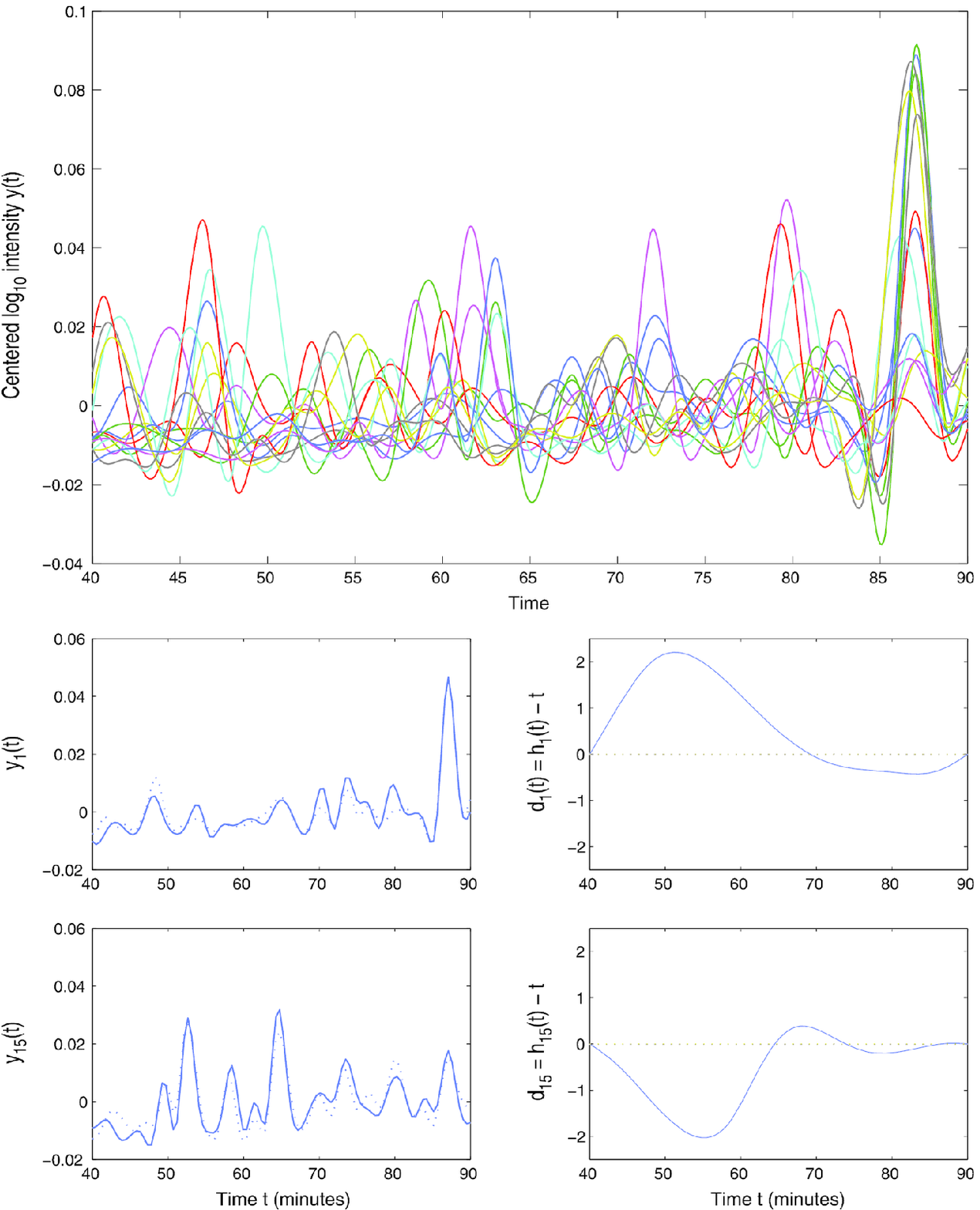}

\caption{Top: Fifteen mean-centered $\log_{10}$-transformations of
sections of mass spectrometry analyses of blood samples.
Bottom left: The fits (dotted curve) to the data (solid curve) for two
observations, $y_1$ and $y_{15}$, produced by three principal
components and registration.
Bottom right: The corresponding warping functions display using
$d_i(t)= h_i(t) -t$.}
\label{ProteomicsData}
\end{figure*}
%
%%%%%%%%%%%%%%%%%%%%%%%%%%%%%%%%%%%%%%%%%%%%%%%%%%%%%%%%%%%%%%%%%%%%%%%%%%%%%%%%%%%%%%%%%%%%%%%%%%%%%%

Shown in Figure~\ref{figwine-data-alignment} is an example of
alignment using the SRVF framework applied
to the wine NMR spectra shown earlier. The top row shows the original
spectra, the aligned spectra, and
the phase functions obtained during the alignment. The bottom row of
Figure~\ref{figwine-data-alignment} shows the same data aligned using
simple shifts and minimizing the loss in (\ref{eqL2-normalized}),
using as a template one of the curves in the sample, the medoid curve,
as detailed, for example, in \citet
{Sangalli-Secchi-Vantini-2014-EJS-AneuRisk-analysis}. For these data
the amplitude variation
is in fact well described by the relative heights of the peaks and the
shifts-warp family is able to capture very well the phase variation in
the part of the spectra here considered, as also highlighted by the
SRVF framework. The associated shifts display a clear clustering in the
phase of the red wines vs the white and ros\'{e} wines. Figure~\ref
{figgrowth-female-data-alignment}
shows the alignment of the growth velocities of the 54 girls in
the Berkeley growth study, in the time interval 3 to 18 years---the
top row displays the results obtained via the SRVF framework, while
the bottom row displays the results obtained
on additionally smoothed data using linear warpings and minimizing the
loss in (\ref{eqL2-normalized}). The nonlinear warping in the SRVF
framework allows for a visibly better alignment of the growth curves,
showing that in many applicative contexts nonlinear warping is indeed
necessary. The linear warping,
obtained after additional smoothing of data, is nevertheless able to
unveil some interesting features of the data. For instance, \citet
{Sangalli-Secchi-Vantini-Vitelli-2010-CSDA} carry out linear warping of
the growth curves of both the girls and boys in the study, highlighting
a neat separation of boys and girls in the phase space and other
interesting aspects of the growth dynamics of the two groups.

Instead of using just one template, it is often beneficial to
divide data into smaller sets and use different templates for alignment
in these subsets. An instance of this
idea is when clustering and alignment are performed together. For
instance, \citet{Sangalli-Secchi-Vantini-Vitelli-2010-CSDA} propose a
$k$-mean alignment procedure that jointly performs alignment and
(unsupervised) clustering of functional data. Other proposals in this
context are given by \citet{Tang-Muller-2009}, \citet{Liu-Yang-2009},
\citet{Boudaoud-Rix-Meste-2010}.
Another set of papers
(\cite{muller-biometrika2008},
\cite{muller-JASA2004},
\cite{Gervini-Gasser2004})
takes the approach where some data points serve as templates for
others, and the
individual warping functions are averaged to find ultimate warpings.

\citet{Kneip-Ramsay-2008} perform registration of functional
observations to the fits
provided by a $K$-dimensional principal components analysis. In other
words, the template is constructed individually for
each function using an orthonormal basis.
As an illustration, consider the 15 sections of mean-centered
log-transformed mass-spectrometry intensities in the top panel of
Figure~\ref{ProteomicsData}. The large peaks
on the right are fairly well registered by a preliminary landmark
registration of the whole sequence, but we see substantial
phase variation in the rest of these spectrum sections that obscures
important amplitude variation.
Three principal components were computed from these data combined with
a registration of each section $y_i$ to its fit $\hat{y}_i$ using a
method currently under development, as well as a principal components
analysis without registration. The mean squared residuals for
unregistered and registered PCA's were 0.0052 and 0.0038, respectively,
corresponding to a squared multiple correlation 0.47. That means nearly
half of the variation around the unregistered fit can be accommodated
by modeling phase variation. The bottom part of Figure~\ref
{ProteomicsData} displays the fits for $y_1$ and $y_{15}$ in its left
panels, along with the deformations $d_i(t) = h_i(t) - t$ associated
with the registration in the right panels. The PCA is able to nicely
accommodate the amplitude variation, and its fits after time warping
are well aligned with all of the peaks. Choice of the number of
components has an important impact on this type of analysis.
Combining registration with model estimation or using multiple
templates further blurs the distinction between amplitude and phase
variation, suggesting that a successful analysis may depend heavily on
prior choices guided by knowledge and intuitions about which type of
variation is the primary focus.

In some contexts it also makes sense to combine the registration
problem with other inferences, such as a regression problem, for a more
comprehensive solution. For instance, \citeauthor{Hadji-etal-arXiv-2013} (\citeyear{Hadji-etal-arXiv-2013,Hadjipantelis-et-al-2014-EJS}) study the problem
of regression using phase and amplitude components
of the given functions.
%%%%%%%%%%%%%%%%%%%%%%%%%%%%%%%%%%%%%%%%%%%%%%%%%%%%%%%%%%%%%%%%%%%%%%%%%%%%%%%%%%%%%%%%%%%%%%%%%%%%%%

%
%%------------------------------------------------------------------------------------------------
%%------------------------------------------------------------------------------------------------

%s4 #&#
\section{Available Software}\label{sec4}
Software implementations of many of the methods illustrated here are
available publicly.
R and Matlab code for implementation of the minimum eigenvalue method
of Ramsay and Silverman can be found at~\surl{http://www.psych.mcgill.ca/misc/fda/downloads/\\FDAfuns/}.
Matlab software for the extended Fisher--Rao SRVF approach of \citet
{Srivastava-et-al-2011-arXiv} is available at
\surl{http://ssamg.stat.fsu.edu/software}
and the R package is available from CRAN under {\it fdasrvf}. The R
package fdakma \citep{fdakma} implementing the k-mean alignment
procedure described in \citet
{Sangalli-Secchi-Vantini-Vitelli-2010-CSDA} is available from CRAN.

%s5 #&#
\section{Discussion and Conclusions}\label{sec5}
\label{secdiscussion}

In this paper we highlight the concept of phase variability that is
present in functional data and the pitfalls of ignoring
it in statistical analysis. After motivating the importance of
phase--amplitude separation, or alignment of functional
data, in statistical analyses we proceed to summarize different ideas
present in the literature for accomplishing this task.
Specifically, we \mbox{describe} the problem of pinching associated with the
classical $\mathbb{L}^2$-norm-based
matching, and present several solutions to avoid this problem. These
solutions involve either restricting
the amount of warping or using an alternative metric to perform matching.

We note that while several methods exist for phase--amplitude
separation, this is not a completely solved problem and forms an active
area of research. A major challenge
comes from the lack of a single mathematical definition or algorithm
that can work in all, or even most, applications and contexts.
For instance, one can argue that the goals of warping in weather data
will be different from that in wine spectra.
Similarly, while in some cases a simple translation and scaling may be
sufficient for alignment of curves,
the other cases require genuine nonlinear warpings for proper alignment.
In some cases effective data analysis is done by seeking the best
possible peak/valley alignment, for example, in spectral data. In those
cases the Fisher--Rao method is the most effective that we have seen so
far. However, in other cases
too much peak alignment can be a distraction, for example, the growth
curve data.
Therefore, it seems more natural to tailor objective functions and
algorithms to the problem area.

Although we have focused on phase--amplitude separation of real-valued
functions in this paper, this problem
is prevalent in several other data object contexts. For instance, the
problem of registration of images is considered
a central issue in medical image registration. See \citet
{sotiras-etal2013} for a recent survey
of warping-based techniques in this problem area. The ideas presented
in this paper can be extended to included
higher-dimensional signals such as images.

%\begin{appendix}
%\section{}
%\end{appendix}

\section*{Acknowledgments}
The authors thank the Mathematical
Biosciences Institute at the
Ohio State University, Columbus, OH, for its support in organizing a workshop
which led to the writing of this paper.

%\begin{supplement}[id=suppA]
%\slink[doi]{10.1214/00-STSXXXXSUPP} %[doi,text={...}] - jei reikia
%suskaldyti doi
%\sfilename{stsXXXX\_supp.pdf}
%\end{supplement}

% imsref loaded by linak, 2015-08-07 10:11:06
%
% imsref loaded by linak, 2015-08-07 15:19:01


\begin{thebibliography}{34}

%b1 ###
\bibitem[\protect\citeauthoryear{Barlow et~al.}{1972}]{barlow1972statistical}
%
\begin{bbook}[mr]
\bauthor{\bsnm{Barlow},~\bfnm{R.~E.}\binits{R.~E.}},
\bauthor{\bsnm{Bartholomew},~\bfnm{D.~J.}\binits{D.~J.}},
\bauthor{\bsnm{Bremner},~\bfnm{J.~M.}\binits{J.~M.}} \AND
\bauthor{\bsnm{Brunk},~\bfnm{H.~D.}\binits{H.~D.}}
(\byear{1972}).
\btitle{Statistical Inference Under Order Restrictions. {T}he Theory
and Application of Isotonic Regression}.
\bpublisher{Wiley},
\blocation{New York}.
\bid{mr={0326887}}
\end{bbook}
%
\iffalse\OrigBibText
%
\begin{bbook}[author]
\bauthor{\bsnm{Barlow},~\bfnm{Richard~E}\binits{R.~E.}},
\bauthor{\bsnm{Bartholomew},~\bfnm{David~J}\binits{D.~J.}},
\bauthor{\bsnm{Bremner},~\bfnm{JM}\binits{J.}} \AND
\bauthor{\bsnm{Brunk},~\bfnm{HD}\binits{H.}}
(\byear{1972}).
\btitle{Statistical inference under order restrictions: The theory and
application of isotonic regression}.
\bpublisher{Wiley New York}.
\end{bbook}
%
\endOrigBibText\fi
\bptok{imsref}%
% NOT OUTPUTTED:
% fpage = xii+388
\endbibitem

%b2 ###
\bibitem[\protect\citeauthoryear{Boudaoud, Rix and
Meste}{2010}]{Boudaoud-Rix-Meste-2010}
%
\begin{barticle}[mr]
\bauthor{\bsnm{Boudaoud},~\bfnm{S.}\binits{S.}},
\bauthor{\bsnm{Rix},~\bfnm{H.}\binits{H.}} \AND
\bauthor{\bsnm{Meste},~\bfnm{O.}\binits{O.}}
(\byear{2010}).
\btitle{Core shape modelling of a set of curves}.
\bjournal{Comput. Statist. Data Anal.}
\bvolume{54}
\bpages{308--325}.
\bid{doi={10.1016/j.csda.2009.08.003}, issn={0167-9473}, mr={2756428}}
\end{barticle}
%
\iffalse\OrigBibText
%
\begin{barticle}[author]
\bauthor{\bsnm{Boudaoud},~\bfnm{S.}\binits{S.}},
\bauthor{\bsnm{Rix},~\bfnm{H.}\binits{H.}} \AND
\bauthor{\bsnm{Meste},~\bfnm{O.}\binits{O.}}
(\byear{2010}).
\btitle{Core Shape modelling of a set of curves}.
\bjournal{Comput. Statist. Data Anal.}
\bvolume{54}
\bpages{308--325}.
\end{barticle}
%
\endOrigBibText\fi
\bptok{imsref}%
% NOT OUTPUTTED:
% number = 2
% doi = http://dx.doi.org/10.1016/j.csda.2009.08.003
% fjournal = Computational Statistics \& Data Analysis
\endbibitem

%b3 ###
\bibitem[\protect\citeauthoryear{Dryden and
Mardia}{1998}]{mardia-dryden-book}
%
\begin{bbook}[mr]
\bauthor{\bsnm{Dryden},~\bfnm{I.~L.}\binits{I.~L.}} \AND
\bauthor{\bsnm{Mardia},~\bfnm{K.~V.}\binits{K.~V.}}
(\byear{1998}).
\btitle{Statistical Shape Analysis}.
\bpublisher{Wiley},
\blocation{Chichester}.
\bid{mr={1646114}}
\end{bbook}
%
\iffalse\OrigBibText
%
\begin{bbook}[author]
\bauthor{\bsnm{Dryden},~\bfnm{I.~L.}\binits{I.~L.}} \AND
\bauthor{\bsnm{Mardia},~\bfnm{K.~V.}\binits{K.~V.}}
(\byear{1998}).
\btitle{Statistical Shape Analysis}.
\bpublisher{John Wiley \& Son}.
\end{bbook}
%
\endOrigBibText\fi
\bptok{imsref}%
% NOT OUTPUTTED:
% isbn = 0-471-95816-6
% fpage = xx+347
\endbibitem

%b4 ###
\bibitem[\protect\citeauthoryear{Gervini and
Gasser}{2004}]{Gervini-Gasser2004}
%
\begin{barticle}[mr]
\bauthor{\bsnm{Gervini},~\bfnm{Daniel}\binits{D.}} \AND
\bauthor{\bsnm{Gasser},~\bfnm{Theo}\binits{T.}}
(\byear{2004}).
\btitle{Self-modelling warping functions}.
\bjournal{J. R. Stat. Soc. Ser. B. Stat. Methodol.}
\bvolume{66}
\bpages{959--971}.
\bid{doi={10.1111/j.1467-9868.2004.B5582.x}, issn={1369-7412}, mr={2102475}}
\end{barticle}
%
\iffalse\OrigBibText
%
\begin{barticle}[author]
\bauthor{\bsnm{Gervini},~\bfnm{D.}\binits{D.}} \AND
\bauthor{\bsnm{Gasser},~\bfnm{T.}\binits{T.}}
(\byear{2004}).
\btitle{Self-modelling warping functions}.
\bjournal{Journal of the Royal Statistical Society, Series B}
\bvolume{66}
\bpages{959--971}.
\end{barticle}
%
\endOrigBibText\fi
\bptok{imsref}%
% NOT OUTPUTTED:
% number = 4
% doi = http://dx.doi.org/10.1111/j.1467-9868.2004.B5582.x
% fjournal = Journal of the Royal Statistical Society. Series B.
%Statistical Methodology
\endbibitem

%b5 ###
\bibitem[\protect\citeauthoryear{Grenander}{1993}]{grenander-patterns-1993}
%
\begin{bbook}[mr]
\bauthor{\bsnm{Grenander},~\bfnm{Ulf}\binits{U.}}
(\byear{1993}).
\btitle{General Pattern Theory}.
\bpublisher{Oxford Univ. Press},
\blocation{New York}.
\bid{mr={1270904}}
\end{bbook}
%
\iffalse\OrigBibText
%
\begin{bbook}[author]
\bauthor{\bsnm{Grenander},~\bfnm{U.}\binits{U.}}
(\byear{1993}).
\btitle{General Pattern Theory}.
\bpublisher{Oxford University Press}.
\end{bbook}
%
\endOrigBibText\fi
\bptok{imsref}%
% NOT OUTPUTTED:
% isbn = 0-19-853671-2
% fpage = xxii+883
\endbibitem

%b6 ###
\bibitem[\protect\citeauthoryear{Grenander and
Miller}{1998}]{grenander-miller98}
%
\begin{barticle}[mr]
\bauthor{\bsnm{Grenander},~\bfnm{Ulf}\binits{U.}} \AND
\bauthor{\bsnm{Miller},~\bfnm{Michael~I.}\binits{M.~I.}}
(\byear{1998}).
\btitle{Computational anatomy: An emerging discipline}.
\bjournal{Quart. Appl. Math.}
\bvolume{56}
\bpages{617--694}.
\bid{issn={0033-569X}, mr={1668732}}
\bptnote{check volume}%
\end{barticle}
%
\iffalse\OrigBibText
%
\begin{barticle}[author]
\bauthor{\bsnm{Grenander},~\bfnm{U.}\binits{U.}} \AND
\bauthor{\bsnm{Miller},~\bfnm{M.~I.}\binits{M.~I.}}
(\byear{1998}).
\btitle{Computational Anatomy: An Emerging Discipline}.
\bjournal{Quarterly of Applied Mathematics}
\bvolume{LVI}
\bpages{617-694}.
\end{barticle}
%
\endOrigBibText\fi
\bptok{imsref}%
% NOT OUTPUTTED:
% number = 4
% coden = QAMAAY
% fjournal = Quarterly of Applied Mathematics
\endbibitem

%b7 ###
\bibitem[\protect\citeauthoryear{Hadjipantelis
et~al.}{2015}]{Hadji-etal-arXiv-2013}
%
\begin{barticle}[mr]
\bauthor{\bsnm{Hadjipantelis},~\bfnm{P.~Z.}\binits{P.~Z.}},
\bauthor{\bsnm{Aston},~\bfnm{J.~A.~D.}\binits{J.~A.~D.}},
\bauthor{\bsnm{M\"uller},~\bfnm{H.~G.}\binits{H.~G.}} \AND
\bauthor{\bsnm{Evans},~\bfnm{J.~P.}\binits{J.~P.}}
(\byear{2015}).
\btitle{Unifying amplitude and phase analysis: A~compositional
data approach to functional multivariate mixed-effects modeling of
Mandarin Chinese}.
\bjournal{J.~Amer. Statist. Assoc.}
\bvolume{110}
\bpages{545--559}.
\bid{mr={3367246}}
\end{barticle}
%
\iffalse\OrigBibText
%
\begin{barticle}[author]
\bauthor{\bsnm{Hadjipantelis},~\bfnm{P.~Z.}\binits{P.~Z.}},
\bauthor{\bsnm{Aston},~\bfnm{J.~A.~D.}\binits{J.~A.~D.}},
\bauthor{\bsnm{M\"uller},~\bfnm{H.~G.}\binits{H.~G.}} \AND
\bauthor{\bsnm{Evans},~\bfnm{J.~P.}\binits{J.~P.}}
(\byear{2013}).
\btitle{Unifying Amplitude and Phase Analysis: A Compositional Data
Approach to
Functional Multivariate Mixed-Effects Modeling of Mandarin Chinese}.
\bjournal{arXiv preprint arXiv:1308.0868}.
\end{barticle}
%
\endOrigBibText\fi
\bptok{imsref}%
\endbibitem

%b8 ###
\bibitem[\protect\citeauthoryear{Hadjipantelis
et~al.}{2014}]{Hadjipantelis-et-al-2014-EJS}
%
\begin{barticle}[mr]
\bauthor{\bsnm{Hadjipantelis},~\bfnm{Pantelis~Z.}\binits{P.~Z.}},
\bauthor{\bsnm{Aston},~\bfnm{John~A.~D.}\binits{J.~A.~D.}},
\bauthor{\bsnm{M\"uller},~\bfnm{Hans-Georg}\binits{H.-G.}} \AND
\bauthor{\bsnm{Moriarty},~\bfnm{John}\binits{J.}}
(\byear{2014}).
\btitle{Analysis of spike train data: A multivariate mixed effects
model for phase and amplitude}.
\bjournal{Electron. J. Stat.}
\bvolume{8}
\bpages{1797--1807}.
\bid{doi={10.1214/14-EJS865E}, issn={1935-7524}, mr={3273597}}
\end{barticle}
%
\iffalse\OrigBibText
%
\begin{barticle}[author]
\bauthor{\bsnm{Hadjipantelis},~\bfnm{P.~Z.}\binits{P.~Z.}},
\bauthor{\bsnm{Aston},~\bfnm{J.~A.~D.}\binits{J.~A.~D.}},
\bauthor{\bsnm{M\"uller},~\bfnm{H.~G.}\binits{H.~G.}} \AND
\bauthor{\bsnm{Moriarty},~\bfnm{J.}\binits{J.}}
(\byear{2014}).
\btitle{Analysis of spike train data: A multivariate mixed effects
model for
phase and amplitude}.
\bjournal{Electron. J. Statist.}
\bvolume{8}
\bpages{1797-1807}.
\end{barticle}
%
\endOrigBibText\fi
\bptok{imsref}%
% NOT OUTPUTTED:
% number = 2
% doi = http://dx.doi.org/10.1214/14-EJS865E
% fjournal = Electronic Journal of Statistics
\endbibitem

%b9 ###
\bibitem[\protect\citeauthoryear{Jung, Dryden and
Marron}{2012}]{jung-etalBiometrika2012}
%
\begin{barticle}[mr]
\bauthor{\bsnm{Jung},~\bfnm{Sungkyu}\binits{S.}},
\bauthor{\bsnm{Dryden},~\bfnm{Ian~L.}\binits{I.~L.}} \AND
\bauthor{\bsnm{Marron},~\bfnm{J.~S.}\binits{J.~S.}}
(\byear{2012}).
\btitle{Analysis of principal nested spheres}.
\bjournal{Biometrika}
\bvolume{99}
\bpages{551--568}.
\bid{doi={10.1093/biomet/ass022}, issn={0006-3444}, mr={2966769}}
\end{barticle}
%
\iffalse\OrigBibText
%
\begin{barticle}[author]
\bauthor{\bsnm{Jung},~\bfnm{S.}\binits{S.}},
\bauthor{\bsnm{Dryden},~\bfnm{I.~L.}\binits{I.~L.}} \AND
\bauthor{\bsnm{Marron},~\bfnm{J.~S.}\binits{J.~S.}}
(\byear{2012}).
\btitle{Analysis of Principle Nested Spheres}.
\bjournal{Biometrika}
\bvolume{99}
\bpages{551-568}.
\end{barticle}
%
\endOrigBibText\fi
\bptok{imsref}%
% NOT OUTPUTTED:
% number = 3
% doi = http://dx.doi.org/10.1093/biomet/ass022
% coden = BIOKAX
% fjournal = Biometrika
\endbibitem

%b10 ###
\bibitem[\protect\citeauthoryear{Kneip and Ramsay}{2008}]{Kneip-Ramsay-2008}
%
\begin{barticle}[mr]
\bauthor{\bsnm{Kneip},~\bfnm{Alois}\binits{A.}} \AND
\bauthor{\bsnm{Ramsay},~\bfnm{James~O.}\binits{J.~O.}}
(\byear{2008}).
\btitle{Combining registration and fitting for functional models}.
\bjournal{J. Amer. Statist. Assoc.}
\bvolume{103}
\bpages{1155--1165}.
\bid{doi={10.1198/016214508000000517}, issn={0162-1459}, mr={2528838}}
\end{barticle}
%
\iffalse\OrigBibText
%
\begin{barticle}[author]
\bauthor{\bsnm{Kneip},~\bfnm{Alois}\binits{A.}} \AND
\bauthor{\bsnm{Ramsay},~\bfnm{James~O.}\binits{J.~O.}}
(\byear{2008}).
\btitle{Combining registration and fitting for functional models}.
\bjournal{Journal of the American Statistical Association}
\bvolume{103}
\bpages{1155--1165}.
\end{barticle}
%
\endOrigBibText\fi
\bptok{imsref}%
% NOT OUTPUTTED:
% number = 483
% doi = http://dx.doi.org/10.1198/016214508000000517
% coden = JSTNAL
% fjournal = Journal of the American Statistical Association
\endbibitem

%b11 ###
\bibitem[\protect\citeauthoryear{Lavine and Workman}{2013}]{Lavine-Workman-2013}
%
\begin{barticle}[pbm]
\bauthor{\bsnm{Lavine},~\bfnm{Barry~K.}\binits{B.~K.}} \AND
\bauthor{\bsnm{Workman},~\bfnm{Jerome. Jr.}\binits{J.~J.}}
(\byear{2013}).
\btitle{Chemometrics}.
\bjournal{Anal. Chem.}
\bvolume{85}
\bpages{705--714}.
\bid{doi={10.1021/ac303193j}, issn={1520-6882}, pmid={23140170}}
\end{barticle}
%
\iffalse\OrigBibText
%
\begin{barticle}[author]
\bauthor{\bsnm{Lavine},~\bfnm{B.~K.}\binits{B.~K.}} \AND
\bauthor{\bsnm{Workman},~\bfnm{J.}\binits{J.} \bsuffix{Jr.}}
(\byear{2013}).
\btitle{Chemometrics}.
\bjournal{Anal. Chem.}
\bvolume{85}
\bpages{705--714}.
\end{barticle}
%
\endOrigBibText\fi
\bptok{imsref}%
% NOT OUTPUTTED:
% number = 2
% fjournal = Analytical chemistry
\endbibitem

%b12 ###
\bibitem[\protect\citeauthoryear{Liu and M\"uller}{2004}]{muller-JASA2004}
%
\begin{barticle}[mr]
\bauthor{\bsnm{Liu},~\bfnm{Xueli}\binits{X.}} \AND
\bauthor{\bsnm{M\"uller},~\bfnm{Hans-Georg}\binits{H.-G.}}
(\byear{2004}).
\btitle{Functional convex averaging and synchronization for
time-warped random curves}.
\bjournal{J. Amer. Statist. Assoc.}
\bvolume{99}
\bpages{687--699}.
\bid{doi={10.1198/016214504000000999}, issn={0162-1459}, mr={2090903}}
\end{barticle}
%
\iffalse\OrigBibText
%
\begin{barticle}[author]
\bauthor{\bsnm{Liu},~\bfnm{X.}\binits{X.}} \AND
\bauthor{\bsnm{M\"uller},~\bfnm{H.~G.}\binits{H.~G.}}
(\byear{2004}).
\btitle{Functional convex averaging and synchronization for
time-warped random
curves}.
\bjournal{J. American Statistical Association}
\bvolume{99}
\bpages{687-699}.
\end{barticle}
%
\endOrigBibText\fi
\bptok{imsref}%
% NOT OUTPUTTED:
% number = 467
% doi = http://dx.doi.org/10.1198/016214504000000999
% coden = JSTNAL
% fjournal = Journal of the American Statistical Association
\endbibitem

%b13 ###
\bibitem[\protect\citeauthoryear{Liu and Yang}{2009}]{Liu-Yang-2009}
%
\begin{barticle}[mr]
\bauthor{\bsnm{Liu},~\bfnm{Xueli}\binits{X.}} \AND
\bauthor{\bsnm{Yang},~\bfnm{Mark~C.~K.}\binits{M.~C.~K.}}
(\byear{2009}).
\btitle{Simultaneous curve registration and clustering for functional data}.
\bjournal{Comput. Statist. Data Anal.}
\bvolume{53}
\bpages{1361--1376}.
\bid{doi={10.1016/j.csda.2008.11.019}, issn={0167-9473}, mr={2657097}}
\end{barticle}
%
\iffalse\OrigBibText
%
\begin{barticle}[author]
\bauthor{\bsnm{Liu},~\bfnm{X.}\binits{X.}} \AND
\bauthor{\bsnm{Yang},~\bfnm{M.~C.~K.}\binits{M.~C.~K.}}
(\byear{2009}).
\btitle{Simultaneous curve registration and clustering for functional data}.
\bjournal{Comput. Statist. Data Anal}
\bvolume{53}
\bpages{1361--1376}.
\end{barticle}
%
\endOrigBibText\fi
\bptok{imsref}%
% NOT OUTPUTTED:
% number = 4
% doi = http://dx.doi.org/10.1016/j.csda.2008.11.019
% fjournal = Computational Statistics \& Data Analysis
\endbibitem

%b14 ###
\bibitem[\protect\citeauthoryear{Lu and Marron}{2013}]{lu2013principal}
%
\begin{bmisc}[author]
\bauthor{\bsnm{Lu},~\bfnm{X.}\binits{X.}} \AND
\bauthor{\bsnm{Marron},~\bfnm{J.~S.}\binits{J.~S.}}
(\byear{2013}).
\bhowpublished{Principal nested spheres for time warped functional
data analysis.
Preprint. Available at \arxivurl{arXiv:1304.6789}.}
\end{bmisc}
%
\iffalse\OrigBibText
%
\begin{barticle}[author]
\bauthor{\bsnm{Lu},~\bfnm{X.}\binits{X.}} \AND
\bauthor{\bsnm{Marron},~\bfnm{J.~S.}\binits{J.~S.}}
(\byear{2013}).
\btitle{Principal Nested Spheres for Time Warped Functional Data Analysis}.
\bjournal{arXiv preprint arXiv:1304.6789}.
\end{barticle}
%
\endOrigBibText\fi
\bptok{imsref}%
\endbibitem

%b15 ###
\bibitem[\protect\citeauthoryear{Marron and
Alonso}{2014}]{Marron-Alonso-2014}
%
\begin{barticle}[mr]
\bauthor{\bsnm{Marron},~\bfnm{J.~Steve}\binits{J.~S.}} \AND
\bauthor{\bsnm{Alonso},~\bfnm{Andr{\'e}s~M.}\binits{A.~M.}}
(\byear{2014}).
\btitle{Overview of object oriented data analysis}.
\bjournal{Biom. J.}
\bvolume{56}
\bpages{732--753}.
\bid{doi={10.1002/bimj.201300072}, issn={0323-3847}, mr={3258083}}
\bptnote{check pages}%
\end{barticle}
%
\iffalse\OrigBibText
%
\begin{barticle}[author]
\bauthor{\bsnm{Marron},~\bfnm{J.~S.}\binits{J.~S.}} \AND
\bauthor{\bsnm{Alonso},~\bfnm{A.~M.}\binits{A.~M.}}
(\byear{2014}).
\btitle{Overview of object oriented data analysis}.
\bjournal{Biometrical Journal}
\bvolume{56}.
\end{barticle}
%
\endOrigBibText\fi
\bptok{imsref}%
% NOT OUTPUTTED:
% number = 5
% doi = http://dx.doi.org/10.1002/bimj.201300072
% fjournal = Biometrical Journal
\endbibitem

%b16 ###
\bibitem[\protect\citeauthoryear{Marron
et~al.}{2014}]{Marron-Ramsay-Sangalli-Srivastava-2014-EJS}
%
\begin{barticle}[mr]
\bauthor{\bsnm{Marron},~\bfnm{J.~S.}\binits{J.~S.}},
\bauthor{\bsnm{Ramsay},~\bfnm{James~O.}\binits{J.~O.}},
\bauthor{\bsnm{Sangalli},~\bfnm{Laura~M.}\binits{L.~M.}} \AND
\bauthor{\bsnm{Srivastava},~\bfnm{Anuj}\binits{A.}}
(\byear{2014}).
\btitle{Statistics of time warpings and phase variations}.
\bjournal{Electron. J. Stat.}
\bvolume{8}
\bpages{1697--1702}.
\bid{doi={10.1214/14-EJS901}, issn={1935-7524}, mr={3273584}}
\end{barticle}
%
\iffalse\OrigBibText
%
\begin{barticle}[author]
\bauthor{\bsnm{Marron},~\bfnm{J.~S.}\binits{J.~S.}},
\bauthor{\bsnm{Ramsay},~\bfnm{J.~O.}\binits{J.~O.}},
\bauthor{\bsnm{Sangalli},~\bfnm{L.~M.}\binits{L.~M.}} \AND
\bauthor{\bsnm{Srivastava},~\bfnm{A.}\binits{A.}}
(\byear{2014}).
\btitle{Statistics of time warpings and phase variations}.
\bjournal{Electron. J. Statist.}
\bvolume{8}
\bpages{1697-1702}.
\end{barticle}
%
\endOrigBibText\fi
\bptok{imsref}%
% NOT OUTPUTTED:
% number = 2
% doi = http://dx.doi.org/10.1214/14-EJS901
% fjournal = Electronic Journal of Statistics
\endbibitem

%b17 ###
\bibitem[\protect\citeauthoryear{Parodi et~al.}{2014}]{fdakma}
%
\begin{bmisc}[author]
\bauthor{\bsnm{Parodi},~\bfnm{Alice}\binits{A.}},
\bauthor{\bsnm{Patriarca},~\bfnm{Mirco}\binits{M.}},
\bauthor{\bsnm{Sangalli},~\bfnm{Laura}\binits{L.}},
\bauthor{\bsnm{Secchi},~\bfnm{Piercesare}\binits{P.}},
\bauthor{\bsnm{Vantini},~\bfnm{Simone}\binits{S.}} \AND
\bauthor{\bsnm{Vitelli},~\bfnm{Valeria}\binits{V.}}
(\byear{2014}).
\bhowpublished{fdakma: Functional data analysis: K-mean alignment.
R package version 1.1.1}.
\end{bmisc}
%
\iffalse\OrigBibText
%
\begin{bmanual}[author]
\bauthor{\bsnm{Parodi},~\bfnm{Alice}\binits{A.}},
\bauthor{\bsnm{Patriarca},~\bfnm{Mirco}\binits{M.}},
\bauthor{\bsnm{Sangalli},~\bfnm{Laura}\binits{L.}},
\bauthor{\bsnm{Secchi},~\bfnm{Piercesare}\binits{P.}},
\bauthor{\bsnm{Vantini},~\bfnm{Simone}\binits{S.}} \AND
\bauthor{\bsnm{Vitelli},~\bfnm{Valeria}\binits{V.}}
(\byear{2014}).
\btitle{fdakma: Functional Data Analysis: K-Mean Alignment}
\bnote{R package version 1.1.1}.
\end{bmanual}
%
\endOrigBibText\fi
\bptok{imsref}%
\endbibitem

%b18 ###
\bibitem[\protect\citeauthoryear{Ramsay and
Silverman}{2005}]{Ramsay-Silverman-2005}
%
\begin{bbook}[mr]
\bauthor{\bsnm{Ramsay},~\bfnm{J.~O.}\binits{J.~O.}} \AND
\bauthor{\bsnm{Silverman},~\bfnm{B.~W.}\binits{B.~W.}}
(\byear{2005}).
\btitle{Functional Data Analysis},
\bedition{2nd} ed.
\bpublisher{Springer},
\blocation{New York}.
\bid{mr={2168993}}
\end{bbook}
%
\iffalse\OrigBibText
%
\begin{bbook}[author]
\bauthor{\bsnm{Ramsay},~\bfnm{J~O}\binits{J.~O.}} \AND
\bauthor{\bsnm{Silverman},~\bfnm{B~W}\binits{B.~W.}}
(\byear{2005}).
\btitle{Functional {D}ata {A}nalysis},
\bedition{Second} ed.
\bpublisher{Springer}.
\end{bbook}
%
\endOrigBibText\fi
\bptok{imsref}%
% NOT OUTPUTTED:
% isbn = 978-0387-40080-8; 0-387-40080-X
% fpage = xx+426
\endbibitem

%b19 ###
\bibitem[\protect\citeauthoryear{Sakoe and Chiba}{1978}]{Sakoe-Chiba-1978}
%
\begin{barticle}[author]
\bauthor{\bsnm{Sakoe},~\bfnm{H.}\binits{H.}} \AND
\bauthor{\bsnm{Chiba},~\bfnm{S.}\binits{S.}}
(\byear{1978}).
\btitle{Dynamic programming algorithm optimization for spoken word
recognition}.
\bjournal{IEEE Trans. Acoust. Speech Signal Process.}
\bvolume{26}
\bpages{43--49}.
\end{barticle}
%
\iffalse\OrigBibText
%
\begin{barticle}[author]
\bauthor{\bsnm{Sakoe},~\bfnm{H.}\binits{H.}} \AND
\bauthor{\bsnm{Chiba},~\bfnm{S.}\binits{S.}}
(\byear{1978}).
\btitle{Dynamic programming algorithm optimization for spoken word
recognition}.
\bjournal{IEEE Transactions on Acoustics, Speech, and Signal Processing}
\bvolume{26}
\bpages{43--49}.
\end{barticle}
%
\endOrigBibText\fi
\bptok{imsref}%
\endbibitem

%b20 ###
\bibitem[\protect\citeauthoryear{Sangalli, Secchi and
Vantini}{2014}]{Sangalli-Secchi-Vantini-2014-EJS-AneuRisk-analysis}
%
\begin{barticle}[mr]
\bauthor{\bsnm{Sangalli},~\bfnm{Laura~M.}\binits{L.~M.}},
\bauthor{\bsnm{Secchi},~\bfnm{Piercesare}\binits{P.}} \AND
\bauthor{\bsnm{Vantini},~\bfnm{Simone}\binits{S.}}
(\byear{2014}).
\btitle{Analysis of {A}neu{R}isk65 data: {$k$}-mean alignment}.
\bjournal{Electron. J. Stat.}
\bvolume{8}
\bpages{1891--1904}.
\bid{doi={10.1214/14-EJS938A}, issn={1935-7524}, mr={3273609}}
\end{barticle}
%
\iffalse\OrigBibText
%
\begin{barticle}[author]
\bauthor{\bsnm{Sangalli},~\bfnm{L.~M.}\binits{L.~M.}},
\bauthor{\bsnm{Secchi},~\bfnm{P.}\binits{P.}} \AND
\bauthor{\bsnm{Vantini},~\bfnm{S.}\binits{S.}}
(\byear{2014}).
\btitle{Analysis of AneuRisk65 data: K-mean Alignment}.
\bjournal{Electron. J. Statist.}
\bvolume{8}
\bpages{1891-1904}.
\end{barticle}
%
\endOrigBibText\fi
\bptok{imsref}%
% NOT OUTPUTTED:
% number = 2
% doi = http://dx.doi.org/10.1214/14-EJS938A
% fjournal = Electronic Journal of Statistics
\endbibitem

%b21 ###
\bibitem[\protect\citeauthoryear{Sangalli
et~al.}{2009}]{Sangalli-Secchi-Vantini-Veneziani-2009-JASA}
%
\begin{barticle}[mr]
\bauthor{\bsnm{Sangalli},~\bfnm{Laura~M.}\binits{L.~M.}},
\bauthor{\bsnm{Secchi},~\bfnm{Piercesare}\binits{P.}},
\bauthor{\bsnm{Vantini},~\bfnm{Simone}\binits{S.}} \AND
\bauthor{\bsnm{Veneziani},~\bfnm{Alessandro}\binits{A.}}
(\byear{2009}).
\btitle{A case study in exploratory functional data analysis:
Geometrical features of the internal carotid artery}.
\bjournal{J. Amer. Statist. Assoc.}
\bvolume{104}
\bpages{37--48}.
\bid{doi={10.1198/jasa.2009.0002}, issn={0162-1459}, mr={2663032}}
\end{barticle}
%
\iffalse\OrigBibText
%
\begin{barticle}[author]
\bauthor{\bsnm{Sangalli},~\bfnm{L.~M.}\binits{L.~M.}},
\bauthor{\bsnm{Secchi},~\bfnm{P.}\binits{P.}},
\bauthor{\bsnm{Vantini},~\bfnm{S.}\binits{S.}} \AND
\bauthor{\bsnm{Veneziani},~\bfnm{A.}\binits{A.}}
(\byear{2009}).
\btitle{A Case Study in Exploratory Functional Data Analysis: Geometrical
Features of the Internal Carotid Artery}.
\bjournal{J. Amer. Statist. Assoc.}
\bvolume{104}
\bpages{37--48}.
\end{barticle}
%
\endOrigBibText\fi
\bptok{imsref}%
% NOT OUTPUTTED:
% number = 485
% doi = http://dx.doi.org/10.1198/jasa.2009.0002
% coden = JSTNAL
% fjournal = Journal of the American Statistical Association
\endbibitem

%b22 ###
\bibitem[\protect\citeauthoryear{Sangalli
et~al.}{2010}]{Sangalli-Secchi-Vantini-Vitelli-2010-CSDA}
%
\begin{barticle}[mr]
\bauthor{\bsnm{Sangalli},~\bfnm{Laura~M.}\binits{L.~M.}},
\bauthor{\bsnm{Secchi},~\bfnm{Piercesare}\binits{P.}},
\bauthor{\bsnm{Vantini},~\bfnm{Simone}\binits{S.}} \AND
\bauthor{\bsnm{Vitelli},~\bfnm{Valeria}\binits{V.}}
(\byear{2010}).
\btitle{{$k$}-mean alignment for curve clustering}.
\bjournal{Comput. Statist. Data Anal.}
\bvolume{54}
\bpages{1219--1233}.
\bid{doi={10.1016/j.csda.2009.12.008}, issn={0167-9473}, mr={2600827}}
\end{barticle}
%
\iffalse\OrigBibText
%
\begin{barticle}[author]
\bauthor{\bsnm{Sangalli},~\bfnm{L.~M.}\binits{L.~M.}},
\bauthor{\bsnm{Secchi},~\bfnm{P.}\binits{P.}},
\bauthor{\bsnm{Vantini},~\bfnm{S.}\binits{S.}} \AND
\bauthor{\bsnm{Vitelli},~\bfnm{V.}\binits{V.}}
(\byear{2010}).
\btitle{K-mean alignment for curve clustering}.
\bjournal{Computational Statistics and Data Analysis}
\bvolume{54}
\bpages{1219--1233}.
\end{barticle}
%
\endOrigBibText\fi
\bptok{imsref}%
% NOT OUTPUTTED:
% number = 5
% doi = http://dx.doi.org/10.1016/j.csda.2009.12.008
% fjournal = Computational Statistics \& Data Analysis
\endbibitem

%b23 ###
\bibitem[\protect\citeauthoryear{Sotiras, Davatzikos and
Paragios}{2013}]{sotiras-etal2013}
%
\begin{barticle}[author]
\bauthor{\bsnm{Sotiras},~\bfnm{A.}\binits{A.}},
\bauthor{\bsnm{Davatzikos},~\bfnm{C.}\binits{C.}} \AND
\bauthor{\bsnm{Paragios},~\bfnm{N.}\binits{N.}}
(\byear{2013}).
\btitle{Deformable medical image registration: A survey}.
\bjournal{IEEE Trans. Med. Imag.}
\bvolume{32}
\bpages{1153--1190}.
\end{barticle}
%
\iffalse\OrigBibText
%
\begin{barticle}[author]
\bauthor{\bsnm{Sotiras},~\bfnm{A.}\binits{A.}},
\bauthor{\bsnm{Davatzikos},~\bfnm{C.}\binits{C.,}} \AND
\bauthor{\bsnm{Paragios},~\bfnm{N.}\binits{N.}}
(\byear{2013}).
\btitle{Deformable Medical Image Registration: A Survey}.
\bjournal{Medical Imaging, IEEE Transactions on}
\bvolume{32}
\bpages{1153-1190}.
\end{barticle}
%
\endOrigBibText\fi
\bptok{imsref}%
\endbibitem




%b24 ###
\bibitem[\protect\citeauthoryear{Srivastava, Jermyn and Joshi}{2007}]{srivastava-etal-Fisher-Rao-CVPR2007}
%
\begin{binproceedings}[author]
\bauthor{\bsnm{Srivastava},~\bfnm{A.}\binits{A.}},
\bauthor{\bsnm{Jermyn},~\bfnm{I.}\binits{I.}} \AND
\bauthor{\bsnm{Joshi},~\bfnm{S.~H.}\binits{S.~H.}}
(\byear{2007}).
\btitle{Riemannian analysis of probability density functions with
applications in vision}.
In \bbooktitle{IEEE Conference on Computer Vision and Pattern Recognition, 2007. CVPR '07}
\bpages{1--8}.
\blocation{Minneapolis, MN, USA}.
\end{binproceedings}
%
\iffalse\OrigBibText
%
\begin{barticle}[author]
\bauthor{\bsnm{Srivastava},~\bfnm{A.}\binits{A.}},
\bauthor{\bsnm{Jermyn},~\bfnm{I.}\binits{I.}} \AND
\bauthor{\bsnm{Joshi},~\bfnm{S.~H.}\binits{S.~H.}}
(\byear{2007}).
\btitle{Riemannian Analysis of Probability Density Functions with Applications
in Vision}.
\bjournal{IEEE Conference on Computer Vision and Pattern Recognition}
\bvolume{0}
\bpages{1-8}.
\end{barticle}
%
\endOrigBibText\fi
\bptok{imsref}%
\endbibitem

%b25 ###
\bibitem[\protect\citeauthoryear{Srivastava
et~al.}{2011a}]{Srivastava-et-al-2011-arXiv}
%
\begin{bmisc}[author]
\bauthor{\bsnm{Srivastava},~\bfnm{A.}\binits{A.}},
\bauthor{\bsnm{Wu},~\bfnm{W.}\binits{W.}},
\bauthor{\bsnm{Kurtek},~\bfnm{S.}\binits{S.}},
\bauthor{\bsnm{Klassen},~\bfnm{E.}\binits{E.}} \AND
\bauthor{\bsnm{Marron},~\bfnm{J.~S.}\binits{J.~S.}}
(\byear{2011}a).
\bhowpublished{Registration of functional data using {F}isher--{R}ao metric.
Preprint. Available at \arxivurl{arXiv:1103.3817v2}.}
\end{bmisc}
%
\iffalse\OrigBibText
%
\begin{barticle}[author]
\bauthor{\bsnm{Srivastava},~\bfnm{A.}\binits{A.}},
\bauthor{\bsnm{Wu},~\bfnm{W.}\binits{W.}},
\bauthor{\bsnm{Kurtek},~\bfnm{S.}\binits{S.}},
\bauthor{\bsnm{Klassen},~\bfnm{E.}\binits{E.}} \AND
\bauthor{\bsnm{Marron},~\bfnm{J.~S.}\binits{J.~S.}}
(\byear{2011}a).
\btitle{Registration of functional data using {F}isher-{R}ao metric}.
\bjournal{arXiv:1103.3817v2 [math.ST]}.
\end{barticle}
%
\endOrigBibText\fi
\bptok{imsref}%
\endbibitem

%b26 ###
\bibitem[\protect\citeauthoryear{Srivastava
et~al.}{2011b}]{srivastavaetalPAMI10}
%
\begin{barticle}[author]
\bauthor{\bsnm{Srivastava},~\bfnm{A.}\binits{A.}},
\bauthor{\bsnm{Klassen},~\bfnm{E.}\binits{E.}},
\bauthor{\bsnm{Joshi},~\bfnm{S.~H.}\binits{S.~H.}} \AND
\bauthor{\bsnm{Jermyn},~\bfnm{I.~H.}\binits{I.~H.}}
(\byear{2011}b).
\btitle{Shape analysis of elastic curves in Euclidean spaces}.
\bjournal{IEEE Trans. Pattern Anal. Mach. Intell.}
\bvolume{33}
\bpages{1415--1428}.
\end{barticle}
%
\iffalse\OrigBibText
%
\begin{barticle}[author]
\bauthor{\bsnm{Srivastava},~\bfnm{A.}\binits{A.}},
\bauthor{\bsnm{Klassen},~\bfnm{E.}\binits{E.}},
\bauthor{\bsnm{Joshi},~\bfnm{S.~H.}\binits{S.~H.}} \AND
\bauthor{\bsnm{Jermyn},~\bfnm{I.~H.}\binits{I.~H.}}
(\byear{2011}b).
\btitle{Shape Analysis of Elastic Curves in Euclidean Spaces}.
\bjournal{IEEE Transactions on Pattern Analysis and Machine Intelligence}
\bvolume{33}
\bpages{1415--1428}.
\end{barticle}
%
\endOrigBibText\fi
\bptok{imsref}%
\endbibitem

%b27 ###
\bibitem[\protect\citeauthoryear{Tang and M{\"
u}ller}{2008}]{muller-biometrika2008}
%
\begin{barticle}[mr]
\bauthor{\bsnm{Tang},~\bfnm{Rong}\binits{R.}} \AND
\bauthor{\bsnm{M{\"u}ller},~\bfnm{Hans-Georg}\binits{H.-G.}}
(\byear{2008}).
\btitle{Pairwise curve synchronization for functional data}.
\bjournal{Biometrika}
\bvolume{95}
\bpages{875--889}.
\bid{doi={10.1093/biomet/asn047}, issn={0006-3444}, mr={2461217}}
\end{barticle}
%
\iffalse\OrigBibText
%
\begin{barticle}[author]
\bauthor{\bsnm{Tang},~\bfnm{R.}\binits{R.}} \AND
\bauthor{\bsnm{M\"{u}ller},~\bfnm{H.~G.}\binits{H.~G.}}
(\byear{2008}).
\btitle{Pairwise curve synchronization for functional data}.
\bjournal{Biometrika}
\bvolume{95}
\bpages{875-889}.
\end{barticle}
%
\endOrigBibText\fi
\bptok{imsref}%
% NOT OUTPUTTED:
% number = 4
% doi = http://dx.doi.org/10.1093/biomet/asn047
% coden = BIOKAX
% fjournal = Biometrika
\endbibitem

%b28 ###
\bibitem[\protect\citeauthoryear{Tang and M{\"
{u}}ller}{2009}]{Tang-Muller-2009}
%
\begin{barticle}[pbm]
\bauthor{\bsnm{Tang},~\bfnm{Rong}\binits{R.}} \AND
\bauthor{\bsnm{M{\"{u}}ller},~\bfnm{Hans-Georg}\binits{H.-G.}}
(\byear{2009}).
\btitle{Time-synchronized clustering of gene expression trajectories}.
\bjournal{Biostatistics}
\bvolume{10}
\bpages{32--45}.
\bid{doi={10.1093/biostatistics/kxn011}, issn={1468-4357},
pii={kxn011}, pmid={18502728}}
\end{barticle}
%
\iffalse\OrigBibText
%
\begin{barticle}[author]
\bauthor{\bsnm{Tang},~\bfnm{R.}\binits{R.}} \AND
\bauthor{\bsnm{M\"{u}ller},~\bfnm{H.~G.}\binits{H.~G.}}
(\byear{2009}).
\btitle{Time-synchronized clustering of gene expression trajectories}.
\bjournal{Biostatistics}
\bvolume{10}
\bpages{32--45}.
\end{barticle}
%
\endOrigBibText\fi
\bptok{imsref}%
% NOT OUTPUTTED:
% number = 1
% fjournal = Biostatistics (Oxford, England)
\endbibitem

%b29 ###
\bibitem[\protect\citeauthoryear{Tucker, Wu and
Srivastava}{2013}]{Tucker-CSDA2013}
%
\begin{barticle}[mr]
\bauthor{\bsnm{Tucker},~\bfnm{J.~Derek}\binits{J.~D.}},
\bauthor{\bsnm{Wu},~\bfnm{Wei}\binits{W.}} \AND
\bauthor{\bsnm{Srivastava},~\bfnm{Anuj}\binits{A.}}
(\byear{2013}).
\btitle{Generative models for functional data using phase and
amplitude separation}.
\bjournal{Comput. Statist. Data Anal.}
\bvolume{61}
\bpages{50--66}.
\bid{doi={10.1016/j.csda.2012.12.001}, issn={0167-9473}, mr={3063000}}
\bptnote{check pages}%
\end{barticle}
%
\iffalse\OrigBibText
%
\begin{barticle}[author]
\bauthor{\bsnm{Tucker},~\bfnm{J.~D.}\binits{J.~D.}},
\bauthor{\bsnm{Wu},~\bfnm{W.}\binits{W.}} \AND
\bauthor{\bsnm{Srivastava},~\bfnm{A.}\binits{A.}}
(\byear{2013}).
\btitle{Generative models for functional data using phase and amplitude
separation}.
\bjournal{Computational Statistics \& Data Analysis}
\bvolume{61}
\bpages{50 - 66}.
\end{barticle}
%
\endOrigBibText\fi
\bptok{imsref}%
% NOT OUTPUTTED:
% doi = http://dx.doi.org/10.1016/j.csda.2012.12.001
% fjournal = Computational Statistics \& Data Analysis
\endbibitem

%b30 ###
\bibitem[\protect\citeauthoryear{Tuddenham and
Snyder}{1954}]{Tuddenham-Snyder-1954}
%
\begin{barticle}[author]
\bauthor{\bsnm{Tuddenham},~\bfnm{R.~D.}\binits{R.~D.}} \AND
\bauthor{\bsnm{Snyder},~\bfnm{M.~M.}\binits{M.~M.}}
(\byear{1954}).
\btitle{Physical growth of California boys and girls from birth to
eighteen years}.
\bjournal{University of California Publication in Child Development}
\bvolume{1}
\bpages{183--364}.
\end{barticle}
%
\iffalse\OrigBibText
%
\begin{barticle}[author]
\bauthor{\bsnm{Tuddenham},~\bfnm{R.~D.}\binits{R.~D.}} \AND
\bauthor{\bsnm{Snyder},~\bfnm{M.~M.}\binits{M.~M.}}
(\byear{1954}).
\btitle{Physical growth of California boys and girls from birth to eighteen
years}.
\bjournal{University of California Publication in Child Development}
\bvolume{1}
\bpages{183--364.}
\end{barticle}
%
\endOrigBibText\fi
\bptok{imsref}%
\endbibitem

%b31 ###
\bibitem[\protect\citeauthoryear{Vantini}{2012}]{Vantini2012}
%
\begin{barticle}[mr]
\bauthor{\bsnm{Vantini},~\bfnm{Simone}\binits{S.}}
(\byear{2012}).
\btitle{On the definition of phase and amplitude variability in
functional data analysis}.
\bjournal{TEST}
\bvolume{21}
\bpages{676--696}.
\bid{doi={10.1007/s11749-011-0268-9}, issn={1133-0686}, mr={2992088}}
\end{barticle}
%
\iffalse\OrigBibText
%
\begin{barticle}[author]
\bauthor{\bsnm{Vantini},~\bfnm{S.}\binits{S.}}
(\byear{2012}).
\btitle{On the definition of phase and amplitude variability in
functional data
analysis}.
\bjournal{TEST}
\bvolume{21}
\bpages{676--696}.
\end{barticle}
%
\endOrigBibText\fi
\bptok{imsref}%
% NOT OUTPUTTED:
% number = 4
% doi = http://dx.doi.org/10.1007/s11749-011-0268-9
% fjournal = TEST
\endbibitem

%b32 ###
\bibitem[\protect\citeauthoryear{Veeraraghavan
et~al.}{2009}]{ashok-srivastava-etal-TIP09}
%
\begin{barticle}[mr]
\bauthor{\bsnm{Veeraraghavan},~\bfnm{Ashok}\binits{A.}},
\bauthor{\bsnm{Srivastava},~\bfnm{Anuj}\binits{A.}},
\bauthor{\bsnm{Roy-Chowdhu\-ry},~\bfnm{Amit~K.}\binits{A.~K.}} \AND
\bauthor{\bsnm{Chellappa},~\bfnm{Rama}\binits{R.}}
(\byear{2009}).
\btitle{Rate-invariant recognition of humans and their activities}.
\bjournal{IEEE Trans. Image Process.}
\bvolume{18}
\bpages{1326--1339}.
\bid{doi={10.1109/TIP.2009.2017143}, issn={1057-7149}, mr={2742162}}
\bptnote{check volume}%
\end{barticle}
%
\iffalse\OrigBibText
%
\begin{barticle}[author]
\bauthor{\bsnm{Veeraraghavan},~\bfnm{A.}\binits{A.}},
\bauthor{\bsnm{Srivastava},~\bfnm{A.}\binits{A.}},
\bauthor{\bsnm{Roy-Chowdhury},~\bfnm{A.~K.}\binits{A.~K.}} \AND
\bauthor{\bsnm{Chellappa},~\bfnm{R.}\binits{R.}}
(\byear{2009}).
\btitle{Rate-invariant Recognition of Humans and Their Activities}.
\bjournal{IEEE Trans. Image Processing}
\bvolume{8}
\bpages{1326-1339}.
\end{barticle}
%
\endOrigBibText\fi
\bptok{imsref}%
% NOT OUTPUTTED:
% number = 6
% doi = http://dx.doi.org/10.1109/TIP.2009.2017143
% coden = IIPRE4
% fjournal = IEEE Transactions on Image Processing
\endbibitem

%b33 ###
\bibitem[\protect\citeauthoryear{Wang and Marron}{2007}]{Wang-Marron-2007}
%
\begin{barticle}[mr]
\bauthor{\bsnm{Wang},~\bfnm{Haonan}\binits{H.}} \AND
\bauthor{\bsnm{Marron},~\bfnm{J.~S.}\binits{J.~S.}}
(\byear{2007}).
\btitle{Object oriented data analysis: Sets of trees}.
\bjournal{Ann. Statist.}
\bvolume{35}
\bpages{1849--1873}.
\bid{doi={10.1214/009053607000000217}, issn={0090-5364}, mr={2363955}}
\end{barticle}
%
\iffalse\OrigBibText
%
\begin{barticle}[author]
\bauthor{\bsnm{Wang},~\bfnm{H.}\binits{H.}} \AND
\bauthor{\bsnm{Marron},~\bfnm{J.~S.}\binits{J.~S.}}
(\byear{2007}).
\btitle{Object oriented data analysis: sets of trees}.
\bjournal{Ann. Statist.}
\bvolume{35}
\bpages{1849--1873}.
\end{barticle}
%
\endOrigBibText\fi
\bptok{imsref}%
% NOT OUTPUTTED:
% number = 5
% doi = http://dx.doi.org/10.1214/009053607000000217
% coden = ASTSC7
% fjournal = The Annals of Statistics
\endbibitem

%b34 ###
\bibitem[\protect\citeauthoryear{Younes
et~al.}{2008}]{younes-michor-mumford-shah08}
%
\begin{barticle}[mr]
\bauthor{\bsnm{Younes},~\bfnm{Laurent}\binits{L.}},
\bauthor{\bsnm{Michor},~\bfnm{Peter~W.}\binits{P.~W.}},
\bauthor{\bsnm{Shah},~\bfnm{Jayant}\binits{J.}} \AND
\bauthor{\bsnm{Mumford},~\bfnm{David}\binits{D.}}
(\byear{2008}).
\btitle{A metric on shape space with explicit geodesics}.
\bjournal{Atti Accad. Naz. Lincei Cl. Sci. Fis. Mat. Natur. Rend.
Lincei (9) Mat. Appl.}
\bvolume{19}
\bpages{25--57}.
\bid{doi={10.4171/RLM/506}, issn={1120-6330}, mr={2383560}}
\end{barticle}
%
\iffalse\OrigBibText
%
\begin{barticle}[author]
\bauthor{\bsnm{Younes},~\bfnm{L.}\binits{L.}},
\bauthor{\bsnm{Michor},~\bfnm{P.~W.}\binits{P.~W.}},
\bauthor{\bsnm{Shah},~\bfnm{J.}\binits{J.}},
\bauthor{\bsnm{Mumford},~\bfnm{D.}\binits{D.}} \AND
\bauthor{\bsnm{Lincei},~\bfnm{R.}\binits{R.}}
(\byear{2008}).
\btitle{A Metric on Shape Space with Explicit Geodesics}.
\bjournal{Matematica E Applicazioni}
\bvolume{19}
\bpages{25--57}.
\end{barticle}
%
\endOrigBibText\fi
\bptok{imsref}%
% NOT OUTPUTTED:
% number = 1
% doi = http://dx.doi.org/10.4171/RLM/506
% fjournal = Atti della Accademia Nazionale dei Lincei. Classe di
%Scienze Fisiche, Matematiche e Naturali. Rendiconti Lincei. Serie IX.
%Matematica e Applicazioni
\endbibitem

\end{thebibliography}
\end{document}